\documentclass[twocolumn,showpacs,showkeys,preprintnumbers,amsmath,amssymb,aps,nofootXinbib,floatfix]{revtex4}
\usepackage{color}
\usepackage{indentfirst}
\usepackage{eufrak}
\usepackage{graphicx}
\usepackage{multirow}
      \def\di{\displaystyle}
      
      \def\bS{{\bf S}}
      \def\bl{{\bf l}}
      \def\bp{{\bf p}}
      
      \def\br{{\bf r}}

      \def\E{{\cal E}}
      
      \def\G{{\cal G}}

      \def\M{{\cal M}}

      \def\P{{\cal P}}
      \def\R{{\cal R}}
      
      \def\T{{\cal T}}
      \def\V{{\cal V}}

      \def\e{{\rm e}}

      \def\u{{{\uparrow\downarrow}}}
      \def\d{{{\downarrow\uparrow}}}
      \def\uu{{{\uparrow\uparrow}}}
      \def\dd{{{\downarrow\downarrow}}}

\begin{document}

\title{Giant Dipole Resonance and Related Spin-dependent Excitations}

\author{ E.B. Balbutsev\email{balbuts@theor.jinr.ru},
I.V. Molodtsova\email{molod@theor.jinr.ru}
}
\affiliation{Bogoliubov Laboratory of Theoretical Physics, Joint Institute for Nuclear Research, 141980 Dubna, Russia}

\begin{abstract}

The time-dependent Hartree–Fock equation is solved by the Wigner Function Moments method
taking into account spin degrees of freedom.
Energies and reduced transition probabilities of $K^\pi=0^-,\ 1^-$ and $2^-$ excitations are calculated taking $^{164}$Dy as an example.
The spin degrees of freedom give rise to the electric Spin Dipole Resonance. Its properties and interplay with the Giant Dipole Resonance are investigated. The deformation-induced splitting of the spin $M2$ resonance is discussed.
The results of calculations are compared with the experimental data and other theoretical studies.

\end{abstract}


\maketitle

\section{Introduction}\label{I}

Collective nuclear dynamics is among the most interesting topics of experimental and theoretical research.
The Wigner Function Moments (WFM) method
is an effective tool for describing collective motion in atomic nuclei (and any other multiparticle systems).
The solution of the Time-Dependent Hartree-Fock-Bogolyubov  equations by the WFM method made it possible to find the energies and excitation probabilities of giant resonances (isoscalar and isovector quadrupole, isoscalar (compression) and isovector dipole) and various low-lying modes.
The nuclear collective motion of the rotational type was studied in the series
of papers \cite{BaSc,Ann,BaMo,BaMoPRC13,BaMoPRC15,BaMoPRC18,BaMoPRC22,BaMoEPJ24}.
It was shown that additionally to the conventional
nuclear scissors mode there can exist the pair of scissors modes connected with the nucleons spin degrees of freedom. They appear due to the spin-orbit part of the nucleus mean field. The conventional scissors mode is generated by the out of phase rotational oscillations of neutrons with respect of protons, whereas two spin modes are generated by the analogous motion of all
"spin-up" nucleons with respect of all "spin-down" nucleons. These three modes are classified as isovector spin-scalar, isovector spin-vector and isoscalar spin-vector ones.
Now it would be quite natural to investigate the nonrotational motions on the
subject of possible existence of the proper spin excitations, for example, the spin giant resonances of various multipolarities.

\begin{figure}[ht!]
\includegraphics[width=0.9\columnwidth]{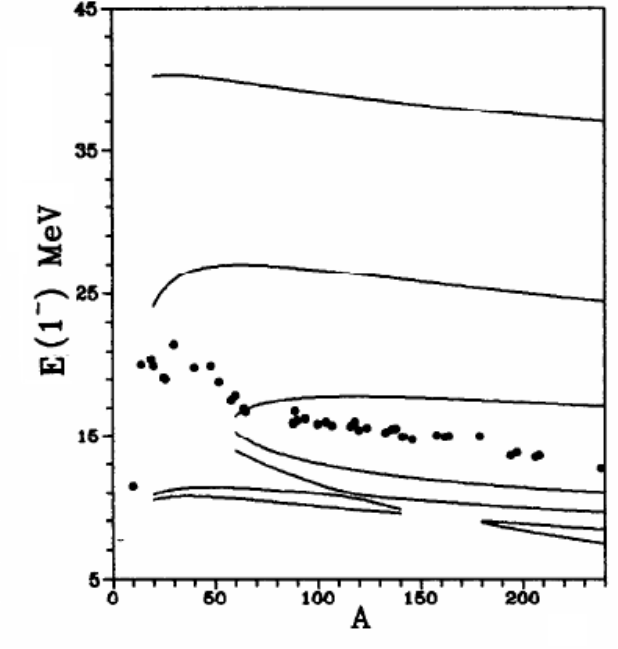}
\caption{Energies of $1^-$ excitations as a function of the mass number calculated in the frame of WFM metod with SkM$^*$ force for nuclei on the beta-stability line. Solid points corresponds to experimental GDR centroids. Reproduced from Ref.~\cite{Piper}.}
\label{fig1}
\end{figure}
\begin{figure}[h!]
\includegraphics[width=0.9\columnwidth]{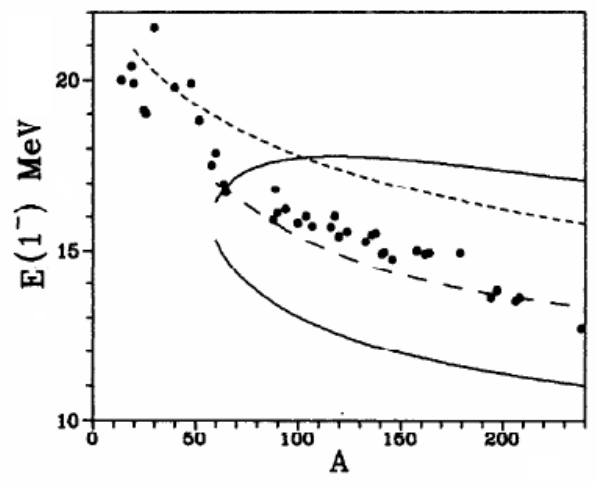}
\caption{The $1^-$ excitations in the GDR energy region.
Calculations taking into account the moments of the first and third ranks. The dashed line  indicates the energy centroid. The dotted line shows the results of calculations taking into account only first-rank tensors. Reproduced from Ref.~\cite{Piper}.}
\label{fig2}
\end{figure}

In this paper we consider the  Giant Dipole Resonance (GDR) and related spin-dependent excitations.
The systematics of the GDR is by now well established~\cite{GDRsyst,Atlas_GDR}.  At the same time, the knowledge about spin-dependent excitations of negative parity
remains rather limited both experimental and theoretical, see~\cite{Morsch78,Sagawa,Castel,Hamamoto,Nester-M2,Paar}.

We already studied $1^-$ excitations in the papers \cite{Piper,Unzha},
where the Time-Dependent Hartree-Fock (TDHF) equation was solved by WFM method
taking into account the moments of first and third ranks.
The calculations were
performed with Skyrme forces neglecting by the deformation of nuclei and by spin degrees of freedom.
Orbital $2^-$ excitations were also studied in this approach~\cite{PiperM2}.

The results for the $1^-$ excitations  are reproduced on Figs.~\ref{fig1},~\ref{fig2}.
The theory produced seven energy levels (Fig.~\ref{fig1}). The centroid of two of
them describes very well the position of GDR energy centroid (Fig.~\ref{fig2}). Two highest levels represent the isoscalar and isovector compressional resonances.
The third from below level represents the toroidal mode.
Two lowest levels can apparently be interpreted as isoscalar and isovector branches of the pygmy resonance.
The simplified calculations, without third rank moments,
produce only one level. It is marked by a dotted line in the Fig.~\ref{fig2}.

Now we are going to perform calculations taking into account spin degrees of freedom and  deformation. The present study is limited by moments of the first rank.
Instead of Skyrme forces the harmonic
oscillator with the spin-orbit potential plus separable dipole-dipole,
quadrupole-quadrupole and spin-dipole--spin-dipole interactions will be used.
The results of calculations for spherical nuclei are reported in Ref.~\cite{BaMoIJMPE26}. We now present an extension of the theory to include deformed nuclei.

In Sec.~\ref{II} we briefly describe the formalism of the WFM method.
The TDHF equations for density matrix are formulated, the model Hamiltonian is presented and collective variables are defined. The derivation of the corresponding dynamical equations is discussed.
Sec.~\ref{III} presents the results of calculations.
In the first, the WFM equations are solved without spin-orbit potential. The ways to avoid the excitation of the center of mass motion are discussed. Further, the exact dynamical equations are introduced and solved. Electric $K^\pi=0^-, 1^-$ and magnetic $K^\pi=0^-, 1^-, 2^-$ excitations of $^{164}$Dy
with corresponding $E1$ and $M2$ strengths are analyzed. The origin and features of the electric spin dipole resonance are studied. The main results are reviewed  and discussed in Sec.~\ref{IV} and conclusions are given in Sec.~\ref{V}.

\section{TDHF equation and WFM equations of motion}\label{II}

 The  TDHF equation reads~\cite{Ring,Solov}
\begin{eqnarray}
i\hbar\dot{\hat\rho} =\hat h\hat\rho -\hat\rho\hat h
\label{HFB}
\end{eqnarray}
Let us consider its matrix form in coordinate
space keeping all spin indices $s, s'$:
$\langle \br,s|\hat\rho|\br',s'\rangle$, etc.
 We do not specify the isospin indices in order to make
formulae more transparent.
After introduction of the more compact notation
$\langle \br,s|\hat X|\br',s'\rangle =X_{rr'}^{ss'}$
the set of equations (\ref{HFB}) with specified spin indices reads
\begin{widetext}
\begin{eqnarray}
&&i\hbar\dot{\rho}_{rr''}^{\uparrow\uparrow} =
\int\!d^3r'\left(
 h_{rr'}^{\uparrow\uparrow}\rho_{r'r''}^{\uparrow\uparrow}
-\rho_{rr'}^{\uparrow\uparrow} h_{r'r''}^{\uparrow\uparrow}
+\hat h_{rr'}^{\uparrow\downarrow}\rho_{r'r''}^{\downarrow\uparrow}
-\rho_{rr'}^{\uparrow\downarrow} h_{r'r''}^{\downarrow\uparrow}
\right),
\nonumber\\
&&i\hbar\dot{\rho}_{rr''}^{\uparrow\downarrow} =
\int\!d^3r'\left(
 h_{rr'}^{\uparrow\uparrow}\rho_{r'r''}^{\uparrow\downarrow}
-\rho_{rr'}^{\uparrow\uparrow} h_{r'r''}^{\uparrow\downarrow}
+\hat h_{rr'}^{\uparrow\downarrow}\rho_{r'r''}^{\downarrow\downarrow}
-\rho_{rr'}^{\uparrow\downarrow} h_{r'r''}^{\downarrow\downarrow}\right),
\nonumber\\
&&i\hbar\dot{\rho}_{rr''}^{\downarrow\uparrow} =
\int\!d^3r'\left(
 h_{rr'}^{\downarrow\uparrow}\rho_{r'r''}^{\uparrow\uparrow}
-\rho_{rr'}^{\downarrow\uparrow} h_{r'r''}^{\uparrow\uparrow}
+\hat h_{rr'}^{\downarrow\downarrow}\rho_{r'r''}^{\downarrow\uparrow}
-\rho_{rr'}^{\downarrow\downarrow} h_{r'r''}^{\downarrow\uparrow}\right),
\nonumber\\
&&i\hbar\dot{\rho}_{rr''}^{\downarrow\downarrow} =
\int\!d^3r'\left(
 h_{rr'}^{\downarrow\uparrow}\rho_{r'r''}^{\uparrow\downarrow}
-\rho_{rr'}^{\downarrow\uparrow} h_{r'r''}^{\uparrow\downarrow}
+\hat h_{rr'}^{\downarrow\downarrow}\rho_{r'r''}^{\downarrow\downarrow}
-\rho_{rr'}^{\downarrow\downarrow} h_{r'r''}^{\downarrow\downarrow}
\right).
\label{HFsp}
\end{eqnarray}
with the conventional notations $\uparrow$ for $s=1/2$ and $\downarrow$ for $s=-1/2$.

We will consider the Wigner transform \cite{Ring} of equations (\ref{HFsp})
(see Ref. \cite{Malov} for mathematical details).
So, instead of four (in $s,s'$) matrix elements of
the density matrix $\rho_{rr'}^{ss'}$
we will work with four Wigner functions $f^{ss'}(\br,\bp)$,
that is more convenient for WFM method
(see Refs. \cite{BaMoPRC13, BaMoPRC15, BaMoPRC18, BaMoPRC22} for details). From now on, we will not write out the coordinate
dependence $(\br,\bp)$ of all functions in order to make formulae
more transparent. We have
\begin{eqnarray}
      i\hbar\dot f^{+} &=&\frac{i\hbar}{2}\{h^+,f^+\}+\frac{i\hbar}{2}\{h^-,f^-\}
+i\hbar\{h^\u ,f^\d\}
+i\hbar\{h^\d ,f^\u\}+...,
\nonumber\\
      i\hbar\dot f^{-} &=&
\frac{i\hbar}{2}\{h^+,f^-\}+\frac{i\hbar}{2}\{h^-,f^+\}
-2h^\d f^\u+2h^\u f^\d
+\frac{\hbar^2}{4}\{\{h^\d,f^\u\}\}
-\frac{\hbar^2}{4}\{\{h^\u,f^\d\}\}+...,
\nonumber\\
      i\hbar\dot f^\u  &=&
f^\u h^-
+\frac{i\hbar}{2}\{h^+,f^\u \}
-\frac{\hbar^2}{8}\{\!\{h^-,f^\u \}\!\}
-h^\u f^-
+\frac{i\hbar}{2}\{h^\u ,f^+\}
+\frac{\hbar^2}{8}\{\!\{h^\u ,f^-\}\!\}+....,
\nonumber\\
      i\hbar\dot f^\d  &=&
-f^\d h^-
+\frac{i\hbar}{2}\{h^+,f^\d \}
+\frac{\hbar^2}{8}\{\!\{h^-,f^\d \}\!\}
+h^\d f^-
+\frac{i\hbar}{2}\{h^\d ,f^+\}
-\frac{\hbar^2}{8}\{\!\{h^\d ,f^-\}\!\}+...,
\label{WHF}
\end{eqnarray}
where the functions $h^{s,s'}$, $f^{s,s'}$
are the Wigner
transforms of $h^{s,s'}_{r,r'}$, $\rho^{s,s'}_{r,r'}$,
respectively,
 $\{f,g\}$ is the Poisson
bracket of the functions $f$ and $g$,
$\{\{f,g\}\}$ is their double Poisson bracket,
$f^{\pm}=f^\uu   {\pm} f^\dd $
and $h^{\pm}=h^\uu   \pm h^\dd $.
The dots stand for terms proportional to higher powers of $\hbar$ -- after
integration over phase space these terms disappear and we arrive to the set of
exact integral equations.

 The microscopic Hamiltonian of the model, harmonic oscillator with
spin-orbit potential plus separable dipole-dipole,
quadrupole-quadrupole and spin-dipole--spin-dipole interactions,
is given by
\begin{eqnarray}
\label{Ham}
 H=\sum\limits_{i=1}^A\left[\frac{\hat\bp_i^2}{2m}+\frac{1}{2}m\omega^2\br_i^2
-\eta\hat \bl_i\hat \bS_i\right]+H_{\rm dd}+H_{\rm qq}+H_{\rm sd},
\end{eqnarray}
with
\begin{eqnarray}
\label{Hdd}
&& H_{\rm dd}=\!
\sum_{\mu=-1}^{1}(-1)^{\mu}
\left\{\bar{\xi}
 \sum\limits_i^Z\!\sum\limits_j^N
+\frac{\xi}{2}
\left[\sum\limits_{i,j(i\neq j)}^{Z}
+\sum\limits_{i,j(i\neq j)}^{N}
\right]
\right\}
r_{-\mu}(i)r_{\mu}(j)
,
\\
\label{Hqq}
&& H_{\rm qq}=\!
\sum_{\mu=-2}^{2}(-1)^{\mu}
\left\{\bar{\kappa}
 \sum\limits_i^Z\!\sum\limits_j^N
+\frac{\kappa}{2}
\left[\sum\limits_{i,j(i\neq j)}^{Z}
+\sum\limits_{i,j(i\neq j)}^{N}
\right]
\right\}
q_{2-\mu}(\br_i)q_{2\mu}(\br_j)
,
\\
\label{Hsd}
&&H_{\rm sd}=\!\sum_{\lambda=0}^{2}
\sum_{\mu=-\lambda}^{\lambda}(-1)^{\mu}
\left\{\bar{\chi}
 \sum\limits_i^Z\!\sum\limits_j^N
+\frac{\chi}{2}
\left[
\sum\limits_{i,j(i\neq j)}^{Z}
+\sum\limits_{i,j(i\neq j)}^{N}
\right]
\right\}
s_{\lambda-\mu}(i) s_{\lambda\mu}(j),
\end{eqnarray}
\end{widetext}
where
$\displaystyle q_{2\mu}(\br)=\sqrt{16\pi/5}\,r^2Y_{2\mu}(\theta,\phi)=
\sqrt{6}\{r\otimes r\}_{2\mu}$,
 $\displaystyle\{r\otimes r\}_{\lambda\mu}=\sum_{\sigma,\nu}
C_{1\sigma,1\nu}^{\lambda\mu}r_{\sigma}r_{\nu},$
$C_{1\sigma,1\nu}^{\lambda\mu}$
is the Clebsch-Gordan coefficient,
 cyclic coordinates $r_{\mu}$ are defined as $r_\mu=\sqrt{{4\pi}/{3}}\,r Y_{1\mu}(\theta,\phi)$~\cite{Var}.
$N$ and $Z$ are the numbers of neutrons and protons, $A=N+Z$, $\xi, \bar{\xi},\kappa,
\bar{\kappa}$, $\chi, \bar{\chi}$ and $\eta$ are strength constants, $s_{\lambda\mu}$ is the spin-dipole operator~\cite{Castel} which is given as
\begin{equation}\label{sd}
 s_{\lambda\mu}=\{\hat S\otimes\hat r\}_{\lambda\mu}/\hbar,
\end{equation}
$\hat S_{\mu}$ are spin matrices~\cite{Var}:
\begin{eqnarray}
 &&\hat S_1=-\frac{\hbar}{\sqrt2}{0\quad 1\choose 0\quad 0},\
\hat S_0=\frac{\hbar}{2}{1\quad\, 0\choose 0\, -\!1},\nonumber\\
 &&\hat S_{-1}=\frac{\hbar}{\sqrt2}{0\quad 0\choose 1\quad 0}.\nonumber
\end{eqnarray}
The mean field generated by $H_{\rm qq}$ was processed in Ref.~\cite{BaMoPRC13}.
The mean fields generated by $H_{\rm dd}$ and $H_{\rm sd}$ are derived and the results are written out in the Appendix~\ref{MeanF}.

Equations (\ref{WHF}) will be solved in a small amplitude
approximation by the WFM method.

\subsection{Dipoles}

Integrating (\ref{WHF}) over phase space with the weights $r_\mu$ and $p_\mu$
one gets dynamical equations for
the following collective variables:
\begin{eqnarray}
&&R^{\tau\zeta}_{\mu}(t)=(2\pi\hbar)^{-3}\!\!\int\!\!  d\br\! \! \int\!\!  d\bp\, r_{\mu}
f^{\tau\zeta}(\br,\bp,t),
\nonumber\\
&&P^{\tau\zeta}_{\mu}(t)=(2\pi\hbar)^{-3}\!\!\int\!\!  d\br\! \! \int\!\!  d\bp\, p_{\mu}
f^{\tau\zeta}(\br,\bp,t),
\label{Varis}
\end{eqnarray}
 where $\tau$  denotes  the isospin index,
\mbox{$\zeta\!=+,\,-,\,\uparrow\downarrow,\,\downarrow\uparrow$}
denotes the spin index: $+$ is spin-scalar (SS), $-$ is spin-vector (SV) and
$\uparrow\downarrow,\,\downarrow\uparrow$ are spin-flip (SF).

The integration yields the sets of coupled (due to neutron-proton interaction)
equations for neutron and proton variables.
The found equations are nonliner due to dipole-dipole,
quadrupole-quadrupole and spin-dipole--spin-dipole interactions.
The small amplitude approximation allows one to linearize the equations.
Writing all variables (\ref{Varis}) as a sum of their equilibrium value (eq) plus a
small deviation
\begin{eqnarray}
&&R^{\tau\zeta}_{\mu}(t)=R^{\tau\zeta}_{\mu}(\mbox{\rm eq})+
\R^{\tau\zeta}_{\mu}(t),
\nonumber\\
&&P^{\tau\zeta}_{\mu}(t)=P^{\tau\zeta}_{\mu}(\mbox{\rm eq})+
\P^{\tau\zeta}_{\mu}(t),
\end{eqnarray}
with
\begin{eqnarray}
&&\R^{\tau\zeta}_{\mu}(t)=(2\pi\hbar)^{-3}\!\!\int\!\!  d\br\! \! \int\!\!  d\bp\, r_{\mu}
\delta f^{\tau\zeta}(\br,\bp,t),
\nonumber\\
&&\P^{\tau\zeta}_{\mu}(t)=(2\pi\hbar)^{-3}\!\!\int\!\!  d\br\! \! \int\!\!  d\bp\, p_{\mu}
\delta f^{\tau\zeta}(\br,\bp,t),
\nonumber
\end{eqnarray}
where $\delta f^{\tau\zeta}$ is a variation of $f^{\tau\zeta}$,
and neglecting quadratic deviations one obtains the linearized equations.
It is convenient to rewrite them
in terms of isovector (IV)
$\bar \R_{\mu}=\R_{\mu}^{\rm n}-\R_{\mu}^{\rm p}$, $\bar \P_{\mu}=\P_{\mu}^{\rm n}-\P_{\mu}^{\rm p}$
and isoscalar (IS)
$\R_{\mu}=\R_{\mu}^{\rm n}+\R_{\mu}^{\rm p}$, $\P_{\mu}=\P_{\mu}^{\rm n}+\P_{\mu}^{\rm p}$
variables.
We also define isovector and isoscalar quadrupole strength constants
$\kappa_1=\frac{1}{2}(\kappa-\bar\kappa)$ and
$\kappa_0=\frac{1}{2}(\kappa+\bar\kappa)$ connected by the relation
$\kappa_1=\alpha\kappa_0$ with $\alpha=-2$~\cite{BaSc}.
In a similar way we introduce isovector and isoscalar dipole
$\xi_1=\frac{1}{2}(\xi-\bar\xi)$, $\xi_0=\frac{1}{2}(\xi+\bar\xi)$ and
spin-dipole $\chi_1=\frac{1}{2}(\chi-\bar\chi)$, $\chi_0=\frac{1}{2}(\chi+\bar\chi)$ strength constants.
The complete set of dynamical equations is written out in the Appendix~\ref{AppEq}.

\section{Results of calculations}\label{III}

First of all it is necessary to determine the values
of the dipole-dipole interaction constants. The most natural way is to reproduce the main characteristics of the GDR,
which is one of the fundamental collective excitation modes in nuclei.
The experimental position of the GDR energy centroid in medium and heavy nuclei can be fairly reproduced by the simple formula,
$E({\rm GDR})\simeq 80 A^{-1/3}$ MeV \cite{Ring}.
It is natural to fix the isovector constant $\xi_1$ in such a way that the theory reproduces the observed excitation energy of GDR.
The fulfillment of this requirement in the collective model developed by Bohr and Mottelson leads to
$ \xi_1=\xi_{\rm BM}=113 A^{-5/3} $ MeV$\cdot$fm$^{-2}$~\cite{BMII}.
The isoscalar constant $\xi_0$ must be chosen so as to
avoid excitation of the Center of Mass Motion (CMM).
To this end, $\xi_0$ will be determined from the
equation of motion for the center of mass of a nucleus.
Following this procedure we avoid the admixture of $E1$ spurious state.

The quadrupole strength constant $\kappa_0$ was determined in Ref.~\cite{Ann}
following the condition of self consistency between the mean field and the
density \cite{BMII}. The choice of spin-dipole strength constants $\chi_1$ and $\chi_0$ will be discussed in the section~\ref{Spin-dependent_eta=0}.

First, let's consider the simplified version of the problem, without the spin-orbit potential ($\eta=0$). In this case, the orbital and spin degrees of freedom are separated.
Such simple model allows us to obtain analytical expressions for the energy and excitation probability of GDR. Comparing them with the results of precise calculations we will estimate the role of spin degrees of freedom.

Further we will analyze the magnetic excitations arising due to spin degrees of freedom.

\subsection{The Giant Dipole Resonance ($\eta=0$ limit)}\label{GDR0}

\subsubsection{ GDR ($K^\pi=1^-$)}\label{sGDRK1}

In the absence of the spin-orbit potential, the system of equations (\ref{mu=1(12)})
splits into three independent subsystems. The subsystems related to spin dynamics will be discussed later.
The subsystem
\begin{widetext}
\begin{eqnarray}\label{mu=1_eta=0}
\dot {\bar\R}^+_1&=&
\frac{1}{m}\bar\P^+_1,
\nonumber\\
\dot {\bar\P}^+_1&=&
-m\tilde\omega^2\left(1+\frac{4}{3}\delta\right)\bar\R^+_1
- \frac{2}{3}\alpha\delta X m\tilde\omega^2\R^+_1
-A\left(\xi_1^{K=1} \bar \R^+_1 + X\xi_0^{K=1} \R^+_1\right),
\nonumber\\
\dot {\R}^+_1&=&
\frac{1}{m}\P^+_1,
\nonumber\\
\dot {\P}^+_1&=&
-m\tilde\omega^2\left(1+\frac{4}{3}\delta\right)\R^+_1
- \frac{2}{3}\alpha\delta X m\tilde\omega^2\bar\R^+_1
-A\left(\xi_0^{K=1}  \R^+_1 + X\xi_1^{K=1} \bar\R^+_1\right)
\end{eqnarray}
describes the coupled dynamics of spin-scalar variables $\bar\R^+_1$, $\bar\P^+_1$, $\R^+_1$, and $\P^+_1$. Here $\delta$ is the deformation parameter,
$\tilde\omega^2=\omega^2/(1+\frac{2}{3}\delta)$.
The isoscalar variables $\R^+_1$ and $\P^+_1$ represent the displacement of the center of mass
and the nucleus momentum variation, respectively.
Isovector variables $\bar\R^+_1$, $\bar\P^+_1$ describe the out of phase translational motion of protons and neutrons (i. e. isovector GDR).
Imposing the time evolution via $\e^{i\Omega t}$ for all variables one transforms (\ref{mu=1_eta=0}) into the set of algebraic equations.
Eigenfrequencies $\Omega^{K=1}$ are found as the solutions of its secular equation:
\begin{eqnarray} \label{GDR_1}
\left[\Omega^{K=1}_\pm\right]^2&=&\tilde\omega^2\left(1+\frac{4}{3}\delta\right)
+\frac{A}{2m}
\left(\xi_0^{K=1}+\xi_1^{K=1}\right)\nonumber\\
&\pm&
\frac{1}{2m}\sqrt{A^2\left(\xi_0^{K=1}-\xi_1^{K=1}\right)^2
+ 4\left(A\xi_0^{K=1}+\frac{2}{3}\delta\alpha m\tilde\omega^2\right)
\left(A\xi_1^{K=1}+\frac{2}{3}\delta\alpha m\tilde\omega^2\right)X^2}.
\end{eqnarray}
Here $ X=(N-Z)/A$ is the asymmetry parameter.
For the rest notations see the Appendix~\ref{AppEq}.

The nucleus translation, i.e. CMM, should be decoupled from the
physical excitations. To ensure this
the isoscalar constant $\xi_0^{K=1}$ is fixed by the condition $\Omega^{K=1}_{\rm IS}\equiv\Omega^{K=1}_{+}=0$. We denote this value
by $\tilde\xi_0^{K=1}$.
From Eq.~(\ref{GDR_1}) we find
\begin{eqnarray} \label{xi0K1}
\tilde\xi_0^{K=1}=-\frac{m\omega^2}{A}\frac{\left(A\xi_1^{K=1}+m\omega^2C_1 \right)C_1 - \frac{2}{3}\alpha\delta\tilde\omega^2/\omega^2
 \left(A\xi_1^{K=1}+\frac{2}{3}\alpha\delta m\tilde\omega^2\right)X^2}{A\xi_1^{K=1}+m\omega^2C_1
 -X^2\left(A\xi_1^{K=1}+\frac{2}{3}\alpha\delta m\tilde\omega^2\right)},
\end{eqnarray}
where the notation $C_1=\left(1+\frac{4}{3}\delta\right)/\left(1+\frac{2}{3}\delta\right)$ is introduced.

For further analysis, it is useful to consider the version of calculations without
coupling between isovector (IV) and isoscalar (IS) systems.
To do this, it is enough to put $X=0$. Separated equations help to classify the obtained solutions.
In this case the formulas (\ref{GDR_1}) and (\ref{xi0K1}) take a particularly simple form:
\begin{eqnarray} \label{GDR_1(0is)}
&&\left[\Omega^{K=1}_{\rm IS}\right]^2
\equiv
\left[\Omega^{K=1}_{+}\right]^2
=\tilde\omega^2\left(1+\frac{4}{3}\delta\right)+\frac{A}{m}\xi_0^{K=1}, \\
\label{GDR_1(0iv)}
&&\left[\Omega^{K=1}_{\rm IV}\right]^2\equiv
\left[\Omega^{K=1}_{-}\right]^2
=\tilde\omega^2\left(1+\frac{4}{3}\delta\right)+\frac{A}{m}\xi_1^{K=1},
\end{eqnarray}
\begin{equation}\label{_xi0K1}
\tilde\xi_0^{K=1}=-\frac{m}{A}\omega^2C_1.
\end{equation}
The same deformation dependence is supposed for the isovector constant:
\begin{equation}\label{xi1K1}
 \xi_1^{K=1}=\xi_{\rm BM}C_1,
\end{equation}
where $\xi_{\rm BM}=113A^{-5/3}$ is taken from~\cite{BMII}.
The numerical estimates in the present work will be given on the example of the nucleus $^{164}$Dy with the deformation parameter $\delta=0.26$.
The values of constants are
\begin{equation}\label{xi01}
 \tilde\xi_0^{K=1}=-0.0126\  {\rm MeV}\!\cdot{\rm fm}^{-2},\quad
\xi_1^{K=1}=0.0264\  {\rm MeV}\!\cdot{\rm fm}^{-2},
\end{equation}
and $\xi_{\rm BM}=0.0230\ {\rm MeV}\!\cdot{\rm fm}^{-2}$. The numerical value of the isoscalar constant given by formula (\ref{xi0K1}) is
$\tilde\xi_0^{K=1}=-0.0130\  {\rm MeV}\!\cdot{\rm fm}^{-2}$.
The comparison of this value with (\ref{xi01}) shows that the influence of isovector-isoscalar coupling is small.

 Excitation probabilities are calculated with the help of the theory of a linear response of a system to a weak external field
 (see Appendix~\ref{AppEM}).
 The reduced probability for an electric dipole transition reads
\begin{equation}\label{BE11pm}
B(E11)_\pm=\pm\, e^2\frac{3\hbar}{4\pi}\frac{NZ}{A}
\frac{m\left(\left[\Omega^{K=1}_\pm\right]^2 -\omega^2 C_1\right) - A\xi_0^{K=1} -
\left(A\xi_1^{K=1}+\frac{2}{3}\delta\alpha m\tilde\omega^2\right)X^2 }
{m\Omega^{K=1}_\pm\sqrt{A^2\left(\xi_0^{K=1}-\xi_1^{K=1}\right)^2
+ 4\left(A\xi_0^{K=1}+\frac{2}{3}\delta\alpha m\tilde\omega^2\right)
\left(A\xi_1^{K=1}+\frac{2}{3}\delta\alpha m\tilde\omega^2\right)X^2}}.
\end{equation}
Neglecting here the isovector-isoscalar coupling and using (\ref{_xi0K1}) and (\ref{GDR_1(0iv)})
we easy obtain:
\begin{eqnarray}
&&B(E11)_{\rm IS}\equiv B(E11)_{+}=0, \\
&&B(E11)_{\rm IV}\equiv B(E11)_{-}= \frac{3}{4\pi}\frac{NZ}{A}\frac{e^2\hbar}{m\Omega_{\rm IV}^{K=1}}.
\label{BE11}
\end{eqnarray}
The results of calculations for $^{164}$Dy are presented in
Table~\ref{tab1}, where the  energies $E_{11}=\hbar\Omega^{K=1}$ of $1^-$ levels with their electric dipole strengths are shown.
\begin{table}[b!]
\caption{The results of WFM calculations for $^{164}$Dy without spin-orbit potential ($\eta=0$ limit):
energies $E_{11}$ and $B(E11)$ strengths of $1^-$ excitations.
The factor 2, which appears due to the degeneration in the sign of $K$, is taken into account for the $E1$ strength.}
\begin{ruledtabular}\begin{tabular}{cccccc}
         \multicolumn{2}{c}{Coupled}  &
        &\multicolumn{3}{c}{Decoupled}\\
 \cline{1-2}   \cline{4-6}
      $E_{11}$, MeV   & $B(E11)\!\!\uparrow$, $e^2$fm$^2$  & &  $E_{11}$, MeV   & $B(E11)\!\!\uparrow$, $e^2$fm$^2$  & \\
  \cline{1-2}   \cline{4-5}
  \multicolumn{5}{c}{$\xi_0^{K=1}/\tilde\xi_0^{K=1}=1$}\\
   ~0.00 & -- & & ~0.00 & ~0.00 & IS \\
   16.28 & 23.86 & & 16.35 & 24.07 & IV \\   \cline{1-2}   \cline{4-5}
 \multicolumn{5}{c}{$\xi_0^{K=1}/\tilde\xi_0^{K=1}=-30$}\\
   52.56 & ~0.00 & & 51.78 & ~0.00 & IS \\
   16.16 & 24.36 & & 16.35 & 24.07 & IV \\
\end{tabular}\end{ruledtabular}
\label{tab1}
\end{table}
The left panel presents the results of calculations with the isosector-isoscalar coupling,
the right panel shows the solutions of decoupled $(X=0)$ equations.
From the comparison of the left and right panels it is clear that the coupling has a minor effect on the energy and excitation probability of GDR.

In the above calculations, we excluded the excitation of CMM by the special choice of the isoscalar constant
$\xi_0^{K=1}=\tilde\xi_0^{K=1}$ (see eq. (\ref{_xi0K1}) ).
However, there is one more possibility to avoid the appearance of the spurious mode. It turns out that the excitation
probability of CMM is quickly decreased with the increase of the absolute value
of $\xi_0^{K=1}$, the GDR energy being practically unchanged.
This situation is also demonstrated by the Table~\ref{tab1}, where the results of calculations for two different values of $\xi_0^{K=1}$ are compared.
In the case of  $\xi_0^{K=1}= \tilde{\xi}_0^{K=1}$ the CMM energy $E_{\rm CM}$ and the corresponding
$E1$ strength are equal to zero (see the first line in the Table~\ref{tab1}),
whereas the GDR energy and its $B(E11)$
have their standard values. In the case of $\xi_0^{K=1}= -30\tilde{\xi}_0^{K=1}$ the energy $E_{\rm CM}$ increase to 52.56 MeV with
$B(E11)\simeq0$, the changes in the  GDR characteristics being very small.
The energies $E_{11}$ of GDR and CMM as the functions of the ratio $\xi_0^{K=1}/\tilde\xi_0^{K=1}$ are shown in Fig.~\ref{fig3}
for $X=0$ (a) and $X\neq0$ (b). One sees that the coupling has little influence on the overall picture of the energy behaviour.

\subsubsection{GDR ($K^\pi=0^-$)}\label{sGDRK0}

In the $\eta=0$ limit, the equations (\ref{mu=0(16)}) for spin-independent variables read
\begin{eqnarray}\label{mu=0_eta=0}
\dot {\bar\R}^+_0&=&
\frac{1}{m}\bar\P^+_0,
\nonumber\\
\dot {\bar\P}^+_0&=&
-m
\tilde\omega^2\left(1-\frac{2}{3}\delta\right)
\bar\R^+_0
+ \frac{4}{3}\alpha\delta X m\tilde\omega^2\R^+_0
-A\left(\xi_1^{K=0} \bar \R^+_0 + X\xi_0^{K=0} \R^+_0\right),
\nonumber\\
\dot {\R}^+_0&=&
\frac{1}{m}\P^+_0,
\nonumber\\
\dot {\P}^+_0&=&-m
\tilde\omega^2\left(1-\frac{2}{3}\delta\right)
\R^+_0
+ \frac{4}{3}\alpha\delta X m\tilde\omega^2\bar\R^+_0
-A\left(\xi_0^{K=0}  \R^+_0 + X\xi_1^{K=0} \bar\R^+_0\right).
\end{eqnarray}

\begin{table*}[b!]
\caption{The results of WFM calculations for $^{164}$Dy without spin-orbit potential ($\eta=0$ limit):
energies $E_{10}$ and $B(E10)$ strengths.
}
\begin{ruledtabular}\begin{tabular}{cccccc}
         \multicolumn{2}{c}{Coupled}  &
        &\multicolumn{3}{c}{Decoupled }\\
 \cline{1-2}   \cline{4-6}
      $E_{10}$, MeV   & $B(E10)$, $e^2$fm$^2$  & &  $E_{10}$, MeV   & $B(E10)$, $e^2$fm$^2$  & \\
  \cline{1-2}   \cline{4-5}
  \multicolumn{5}{c}{$\xi_0^{K=0}/\tilde\xi_0^{K=0}=1$}\\
   ~0.00 & -- & & ~0.00 & ~0.00 & IS \\
   12.80 & 14.85 & & 12.81 & 15.36 & IV \\ \cline{1-2}   \cline{4-5}
 \multicolumn{5}{c}{$\xi_0^{K=0}/\tilde\xi_0^{K=0}=-30$}\\
   40.76 & ~0.00 & & 40.57 & ~0.00 & IS \\
   12.56 & 15.67 & & 12.81 & 15.36 & IV \\
\end{tabular}\end{ruledtabular}
\label{tab2}
\end{table*}

\begin{figure*}[t!]
\includegraphics[width=0.4\columnwidth]{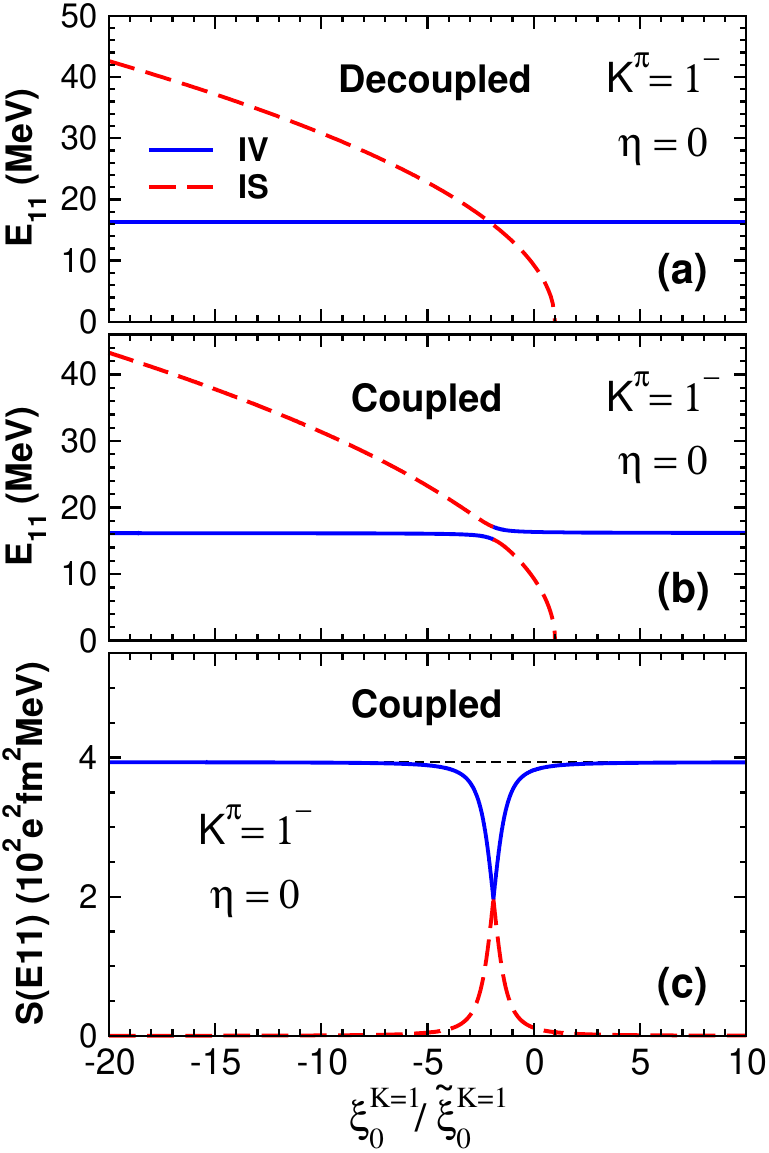}\qquad\includegraphics[width=0.4\columnwidth]{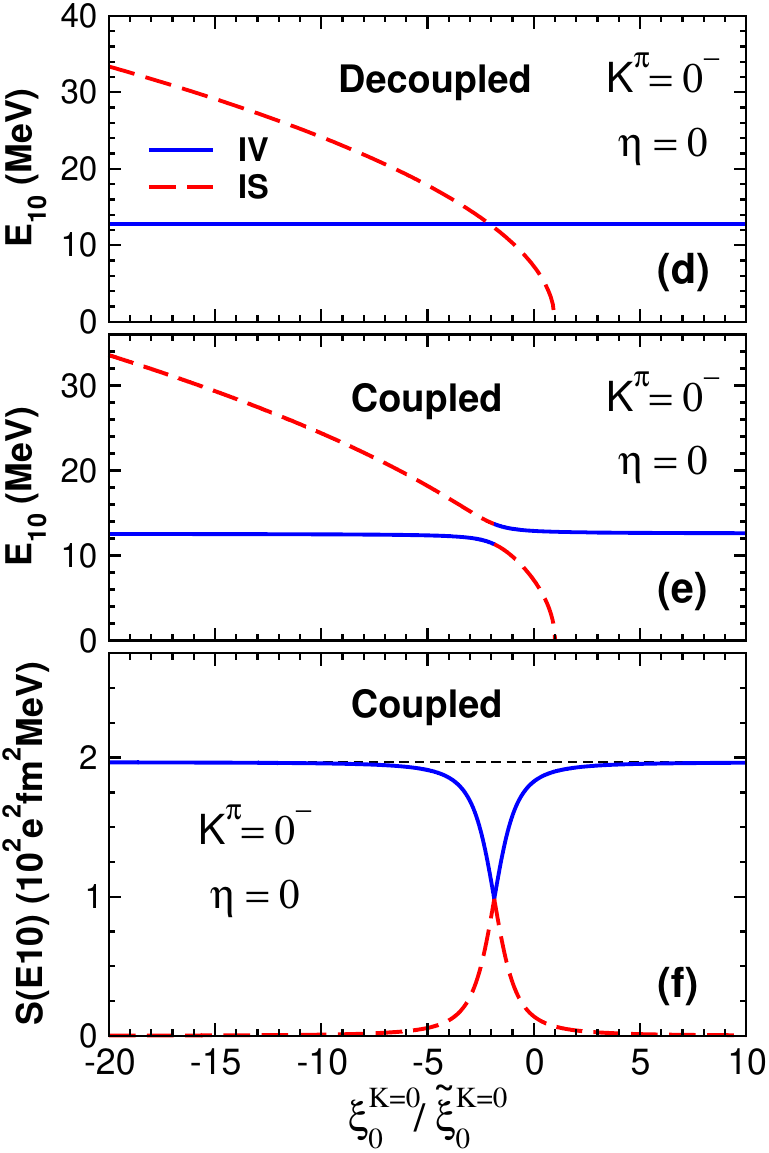}
\caption{Energies $E_{1K}$ (a, b, d, e) and values of $S(E1K)=E_{1K}B(E1K)$ (c, f) vs. $\xi_0^{K=\mu}/\tilde\xi_0^{K=\mu}$ ratio.
The calculations are performed for $^{164}$Dy without the spin-orbit interaction ($\eta=0$).
The solid blue line corresponds to the GDR, and the dotted red line to the CMM.
The black short dashed lines in the panels (c) and (f) indicate the EWSR (\ref{SRE1}) values.}
\label{fig3}
\end{figure*}
The characteristic equation of the respective set of algebraic equations has
two eigenvalues
\begin{eqnarray}\label{GDR_0}
\left[\Omega^{K=0}_\pm\right]^2&=&\tilde\omega^2\left(1-\frac{2}{3}\delta\right)
+\frac{A}{2m}
\left(\xi_0^{K=0}+\xi_1^{K=0}\right)\nonumber \\
&\pm&
\frac{1}{2m}\sqrt{A^2\left(\xi_0^{K=0}-\xi_1^{K=0}\right)^2
+ 4\left(A\xi_0^{K=0}-\frac{4}{3}\delta\alpha m\tilde\omega^2\right)
\left(A\xi_1^{K=0}-\frac{4}{3}\delta\alpha m\tilde\omega^2\right)X^2}.
\end{eqnarray}
Without isovector-isoscalar coupling ($X=0$) they are simplified to:
\begin{eqnarray} \label{GDR_0(0is)}
&&\left[\Omega^{K=0}_{\rm IS}\right]^2\equiv\left[\Omega^{K=0}_{+}\right]^2=
\tilde\omega^2\left(1-\frac{2}{3}\delta\right)
+\frac{A}{m}\xi_0^{K=0},\\
\label{GDR_0(0iv)}
&&\left[\Omega^{K=0}_{\rm IV}\right]^2\equiv\left[\Omega^{K=0}_{-}\right]^2=
\tilde\omega^2\left(1-\frac{2}{3}\delta\right)
+\frac{A}{m}\xi_1^{K=0}.
\end{eqnarray}
Analogously to the case $K^\pi=1^-$ we find from the condition
$\Omega^{K=0}_{\rm IS}=0$:
\begin{equation}\label{_xi0K0}
\tilde\xi_0^{K=0}=-\frac{m}{A}\omega^2C_0,
\end{equation}
where $C_0=\left(1-\frac{2}{3}\delta\right)/\left(1+\frac{2}{3}\delta\right)$.
For the isovector constant in~(\ref{GDR_0(0iv)}) we take
\begin{equation} \label{x11K0}
\xi_1^{K=0}=\xi_{\rm BM} C_0,
\end{equation}
Numerical values of these constants for $^{164}$Dy are
\begin{equation}\label{xi00}
 \tilde\xi_0^{K=0}=-0.0077\  {\rm MeV}\!\cdot{\rm fm}^{-2},\quad
 \xi_1^{K=0}=0.0162\  {\rm MeV}\!\cdot{\rm fm}^{-2}.
\end{equation}
In general case ($X\ne 0$) the condition  $\Omega^{K=0}_+=0$ is satisfied by
\begin{eqnarray} \label{xi0K0}
\tilde\xi_0^{K=0}=-\frac{m}{A}\frac{\left(A\xi_1^{K=0}+m\omega^2C_0 \right)
\omega^2C_0 + \frac{4}{3}\alpha\delta\tilde\omega^2
 \left(A\xi_1^{K=0}-\frac{4}{3}\alpha\delta m\tilde\omega^2\right)X^2}{A\xi_1^{K=0}+m\omega^2C_0
 -X^2\left(A\xi_1^{K=0}-\frac{4}{3}\alpha\delta m\tilde\omega^2\right)}.
\end{eqnarray}
For $^{164}$Dy eq. (\ref{xi0K0}) gives
$\tilde\xi_0^{K=0}=-0.0078\  {\rm MeV}\!\cdot{\rm fm}^{-2}$,
i.e. the influence of isovector-isoscalar coupling on the value of the $\tilde\xi_0^{K=0}$ is very small (exactly as in the case $K^\pi=1^-$).
The excitation probability is calculated in the same way as for $K^\pi=1^-$.
The $B(E10)$ strength reads
\begin{equation}\label{BE10pm}
B(E10)_\pm=\pm\, e^2\frac{3\hbar}{4\pi}\frac{NZ}{A}
\frac{m\left(\left[\Omega^{K=0}_\pm\right]^2 -\omega^2 C_0\right) - A\xi_0^{K=0} -
\left(A\xi_1^{K=0}-\frac{4}{3}\delta\alpha m\tilde\omega^2\right)X^2 }
{m\Omega^{K=0}_\pm\sqrt{A^2\left(\xi_0^{K=0}-\xi_1^{K=0}\right)^2
+ 4\left(A\xi_0^{K=0}-\frac{4}{3}\delta\alpha m\tilde\omega^2\right)
\left(A\xi_1^{K=0}-\frac{4}{3}\delta\alpha m\tilde\omega^2\right)X^2}}.
\end{equation}
Neglecting the terms proportional to $X$ one gets
\begin{eqnarray}
&&B(E10)_{\rm IS}\equiv B(E10)_{+}=0, \\
&&B(E10)_{\rm IV}\equiv B(E10)_{-}= \frac{3}{4\pi}\frac{NZ}{A}\frac{e^2\hbar}{2m\Omega_{\rm IV}^{K=0}}.
\label{BE10}
\end{eqnarray}

\end{widetext}

The results of calculations for
$^{164}$Dy are presented in the Table~\ref{tab2}.
We again have considered various scenarios for eliminating CMM excitation.
The results are given for two different values of $\xi_0^{K=0}$.
The energies $E_{10}$ of GDR and CMM excitations as functions of the
$\xi_0^{K=0}/\tilde\xi_0^{K=0}$ are shown in Fig.~\ref{fig3}
for $X=0$ (d) and $X\neq0$ (e). The influence of the isovector-isoscalar coupling is insignificant analogously to the case of $K^\pi=1^-$.

\subsubsection{GDR sum rule}

The experimental position of the GDR energy centroid
in medium and heavy nuclei follows a simple $A$ dependence:
$E_{\rm GDR}=80A^{-1/3}$ [MeV].
In  deformed  axially symmetric nuclei, the GDR strength is split into two components, corresponding to oscillations of neutrons against protons
along ($K=0$) and perpendicular ($|K|= 1$) to the symmetry axis.
The energy separation was found to be proportional to the ground state deformation $\delta$:
$\Delta_{\rm GDR}=E_{11}-E_{10}\simeq E_{\rm GDR}\delta$~\cite{Suzuki77}.
For $^{164}$Dy, the above fitting formulas produce
$E_{\rm GDR}=14.62$~MeV and $\Delta_{\rm GDR}=3.80$~MeV.
Our calculations within the WFM method lead to the energy position $E_{10}=12.80\ (12.56)$~MeV
and $E_{11}=16.28\ (16.16)$~MeV
(see Tables~\ref{tab1} and~\ref{tab2}). So for
the GDR centroid and splitting we find $\bar E_{\rm GDR}=14.94\ (14.75)$ MeV and
$\Delta_{\rm GDR}=3.48\ (3.60)$ MeV.
These values are in fairly good agreement with the empirical estimates for GDR given above.

We find that our calculations exhaust 100\% of the classical energy-weighted sum rule (EWSR) \cite{BMII}
\begin{equation}\label{SRE1}
S_1(E1,K)= \frac{3}{4\pi}\frac{\hbar^2}{2m}\frac{NZ}{A}e^2,
\end{equation}
when the energy of the CMM is pushed to the high energy region.
If the admixture of $E1$ spurious state is eliminated by zeroing the
CMM excitation energy, then the exhaustion of the sum rule
is 98\%.
The rest 2\% are taken by CMM.
The results of calculations of $S(E1K)=E_{1K}B(E1K)$ in the case of $X\neq0$ (coupled) are shown in Figs.~\ref{fig3} (c) and (f)
depending on the ratio $\xi_0^{K}/\tilde\xi_0^{K}$ for $^{164}$Dy.
The black tiny dashed lines indicate the EWSR values given by Eq.~(\ref{SRE1}).

The Figs.~\ref{fig3} (b,e) demonstrate the usual quantum mechanical quasi-crossings of the energy levels.
The reliable results for the GDR energy and the sum rule are obtained for all $\xi_0$
excluding the values in the vicinity of the quasi-crossing point.

\subsection{Spin-dependent excitations ($\eta=0$ limit)}
\label{Spin-dependent_eta=0}

In this paper, the study is limited by tensors of the first rank.
As follows from~(\ref{M2sl}), the orbital part of the $M2$ response is expressed through third-rank tensors~(\ref{T3}).
Orbital $2^-$ excitations were studied within the WFM method using Skyrme forces in the paper~\cite{PiperM2}.
However, microscopic estimates indicate that $M2$ transitions are mainly determined by the spin response, see~\cite{JdePhys,Zr-exp2,Paar}.
In the present consideration the spin contribution~(\ref{M2s}) to $B(M2)$ will be estimated.

\subsubsection{$K^\pi=1^-$}

In the $\eta=0$ limit, the system of equations (\ref{mu=1(12)}) splits into three independent subsets.
The one that describes the $K^\pi=1^-$ component of GDR (\ref{mu=1_eta=0}), was analyzed above.
The second subsystem describes the dynamics of SV isovector $\bar\R^-_1$, $\bar\P^-_1$ and SV isoscalar $\R^-_1$, $\P^-_1$ variables:
\begin{eqnarray}
\dot {\bar\R}^-_1&=&
\frac{1}{m}\bar\P^-_1,
\nonumber \\
\dot {\bar\P}^-_1&=&
-m\omega^2C_1\bar\R^-_1  - \frac{2}{3}\alpha\delta Xm\tilde\omega^2\R^-_1
\nonumber\\&&
-\frac{A}{4}\left(\chi_1 \bar\R^-_1 + \chi_0 X\R^-_1\right)
\nonumber \\
\dot {\R}^-_1&=&
\frac{1}{m}\P^-_1,
\nonumber \\
\dot {\P}^-_1&=&
-m\omega^2C_1\R^-_1   - \frac{2}{3}\alpha\delta Xm\tilde\omega^2\bar\R^-_1
\nonumber\\&&
-\frac{A}{4}\left(\chi_0 \R^-_1 + \chi_1 X\bar\R^-_1\right).
\label{K=1sv(eta=0)}\end{eqnarray}
The solution of its  secular equation gives:
\begin{eqnarray} \label{sv_M21(eta=0)}
\left[\Omega^{K=1}_\pm\right]^2_{\rm SV}=\omega^2C_1+\frac{A}{8m}
\left(\chi_0+\chi_1\pm\sqrt{\G_1}\right),
\end{eqnarray}
where
\begin{eqnarray}
&&\G_1=\left(\chi_0-\chi_1\right)^2\nonumber\\
&&+ 4X^2\left[\chi_0+\frac{8}{3A}\delta\alpha m\tilde\omega^2\right]
\left[\chi_1+\frac{8}{3A}\delta\alpha m\tilde\omega^2\right].\quad
\end{eqnarray}
The third subsystem contains only SF variables:
\begin{eqnarray}
\dot {\bar\R}^\d_0&=&
\frac{1}{m}\bar\P^\d_0,
\nonumber \\
\dot {\bar\P}^\d_0&=&
-m\omega^2C_0\bar\R^\d_0 + \frac{4}{3}\alpha\delta Xm\tilde\omega^2\R^\d_0
\nonumber\\&&
-\frac{A}{4}\left(\chi_1 \bar\R^\d_0 + \chi_0 X\R^\d_0\right),
\nonumber \\
\dot {\R}^\d_0&=&
\frac{1}{m}\P^\d_0,
\nonumber \\
\dot {\P}^\d_0&=&
-m\omega^2C_0\R^\d_0   + \frac{4}{3}\alpha\delta Xm\tilde\omega^2\bar\R^\d_0
\nonumber\\&&
-\frac{A}{4}\left(\chi_0 \R^\d_0 + \chi_1 X\bar\R^\d_0\right).
\label{K=1sf(eta=0)}\end{eqnarray}
The corresponding eigenfrequencies are
\begin{eqnarray} \label{sf_M21(eta=0)}
\left[\Omega^{K=1}_\pm\right]^2_{\rm SF}=\omega^2C_0+\frac{A}{8m}
\left(\chi_0+\chi_1\pm\sqrt{\G_0}\right),
\end{eqnarray}
with
\begin{eqnarray}
&&\G_0=\left(\chi_0-\chi_1\right)^2\nonumber\\
&&+ 4X^2\left[\chi_0-\frac{16}{3A}\delta\alpha m\tilde\omega^2\right]
\left[\chi_1-\frac{16}{3A}\delta\alpha m\tilde\omega^2\right].\quad
\end{eqnarray}
As follows from equation (\ref{M21}), the above variables are related to the $K^\pi=1^-$ branch of the spin magnetic quadrupole resonance.

\begin{table}[h!]
\caption{The magnetic spin-vector (SV) and spin-flip (SF) $K^\pi=1^-$ excitations of $^{164}$Dy.}
\begin{ruledtabular}\begin{tabular}{cccccccc}
 & \multicolumn{2}{c}{Coupled}  &
        &\multicolumn{3}{c}{Decoupled}\\
        \cline{2-3}   \cline{5-7}
  &         $E_{21}^{\rm SF}$   & $B(M21)\!\!\uparrow$  & &  $E_{21}^{\rm SF}$   & $B(M21)\!\!\uparrow$ &  \\
\cline{2-3}   \cline{5-7}
 \multirow{2}{*}{SV} & ~7.61 & ~~14.89 & & ~7.55  & ~102.72 & IS \\
                     &12.02 & 1766.92 & & 12.06  & 1841.27 & IV \\
\hline
 \multirow{2}{*}{SF} &  ~4.79 & ~212.62 & & ~4.86  & ~159.57 & IS \\
                     & 10.61 & 1916.18 & & ~10.58 & 2097.89 & IV \\
\end{tabular}\end{ruledtabular}
\label{tab3}
\end{table}
The energy position of the $M2$ states depends on the spin-dipole interaction constants $\chi_0$ and $\chi_1$. We adopt the $\di\chi_0=-\frac{4\pi\kappa_{\rm sd}}{A \langle r^2\rangle}\frac{3}{4\pi}$ MeV/fm$^2$ with $\chi_1=\alpha_1\chi_0$.
The value $\kappa_{\rm sd}= 25$ Mev was chosen by analyzing the $2^-$ strength distribution in $^{208}$Pb within the RPA in Ref.~\cite{Hamamoto}. The results of our calculations using this value of $\kappa_{\rm sd}$ for $^{208}$Pb~\cite{BaMoIJMPE26}  are in overall agreement with the results obtained within the RQRPA framework, see Ref.~\cite{Paar}.
Since there are no other justified criteria for fixing the value of this constant, we adopt $\kappa_{\rm sd}=25$ MeV for our calculations. We also take  $\alpha_1=-2$,
the same as for the quadrupole-quadrupole interaction.

The calculated energies and $B(M21)\!\!\uparrow$ values for $^{164}$Dy are shown in the  Table~\ref{tab3}.
 The quenching $q=0.7$ of the spin gyromagnetic factor ($g_s=q g_s^{\rm free}$) was used in the calculations of magnetic strengths, see Appendix~\ref{AppB}.
Most of the strength is contained in the isovector excitations. The summed value of $B(M21)\!\!\uparrow$ for all four $1^-$ states is 3910.60 $\mu_{N}^2$fm$^2$.

 The EWSR $S_1^\sigma(M2,K)$ coming from double commutator of the spin part of $\hat O(M2,K)$ operator (\ref{OM2sl}) with the Hamiltonian takes the following form:
 \begin{eqnarray}
 S^\sigma(M2,K)= \frac{15}{16\pi}\frac{\hbar^2}{m}\left[(g_s^{\rm n})^2N + (g_s^{\rm p})^2Z\right]\mu_N^2.\quad
 \label{SR_M20}
 \end{eqnarray}
 It is easy to check that energy-weighted summation over the states listed in the  Tables~\ref{tab3}
 exhausts 100\% of the EWSR (\ref{SR_M20}).

\subsubsection{$K^\pi=0^-$}\label{K=0,eta=0}

By the simple change of variables
\begin{eqnarray}\label{RP0E}
&&\R_0^E=\R^\u_1+\R^\d_{-1},\quad
\P_0^E=\P^\u_1+\P^\d_{-1}\\
\label{RP0M}
&&\R_0^M=\R^\u_1-\R^\d_{-1},\quad
\P_0^M=\P^\u_1-\P^\d_{-1},
\end{eqnarray}
the system (\ref{mu=0(16)}) of dynamical equations
is transformed into three independent subsystems (\ref{mu=0(E)}),  (\ref{mu=0(M-SV)}) and (\ref{mu=0(M-SF)}).

In the case $\eta=0$ the set (\ref{mu=0(E)}) in its turn splits into two  independent subsets.
The one describes the $K^\pi=0^-$ branch of GDR, see eq.~(\ref{mu=0_eta=0}).
The remaining subsystem for the variables $\R_0^E$, $\P_0^E$ reads:
\begin{eqnarray}
\dot {\bar\R}_0^E&=&
\frac{1}{m}\bar\P_0^E,
\nonumber \\
\dot {\bar\P}_0^E&=&
-m\omega^2C_1\bar\R_0^E  - \frac{2}{3}\alpha\delta Xm\tilde\omega^2\R_0^E
\nonumber\\&&
-\frac{A}{4}\left(\chi_1 \bar\R_0^E + \chi_0 X\R_0^E\right)
\nonumber \\
\dot {\R}_0^E&=&
\frac{1}{m}\P_0^E,
\nonumber \\
\dot {\P}_0^E&=&
-m\omega^2C_1\R_0^E   - \frac{2}{3}\alpha\delta Xm\tilde\omega^2\bar\R_0^E
\nonumber\\&&
-\frac{A}{4}\left(\chi_0 \R_0^E + \chi_1 X\bar\R_0^E\right).
\label{SD10(eta=0)}\end{eqnarray}
The eqs.~(\ref{SD10(eta=0)}) formally coincide with eq.~(\ref{K=1sv(eta=0)}) --
the only difference is in the notations of variables. Naturally the eigenfrequencies of both systems also coincide.
The respective two states in $^{164}$Dy have energies 7.61 MeV and 12.02 MeV.
However they are not excited, because the variables $\R_0^E, \P_0^E$ are not
disturbed by any of the electromagnetic operators, see Appendix~\ref{AppEM}.
The situation is changed in the case $\eta\ne 0$ (see Table~\ref{tab4}).

The sets (\ref{mu=0(M-SV)}) and (\ref{mu=0(M-SF)})
describe spin magnetic $K^\pi=0^-$ transitions.
It is necessary to emphasize that separation
occurs already in the general case (i.e. for $\eta\ne 0$).
Nevertheless even in this case the sets (\ref{mu=0(M-SV)}) and (\ref{mu=0(M-SF)}) do not include $\eta$. That is why it will be analyzed in the Section~\ref{M2K0}.


\subsection{Exact calculations ($\eta\neq 0$)}

Now we repeat the calculations, this time taking into account the spin-orbit potential. In this case, the $\tilde\xi_0^K$ constant of the dipole-dipole interaction reads
\begin{equation}\label{xi_K}
\tilde\xi_0^K=\frac{-\xi_1^K A\left(B_K-X^2D_K\right)-B_K^2+X^2D_K^2}{A\left(\xi_1^K A(1-X^2)+B_K-X^2D_K\right)},
\end{equation}
where $\xi_1^K=\xi_{\rm BM}C_K$, $\di B_K=m\omega^2C_K-\frac{\hbar^2}{4}\eta^2$ and $\di D_1=\frac{2}{3} m\tilde\omega^2\delta\alpha$, $\di D_0=-\frac{4}{3} m\tilde\omega^2\delta\alpha$.
The constants of the spin-dipole--spin-dipole interaction $\chi_0$ and $\chi_1$ are the same as in the previous section.

The found states are of a complicated nature.
However, based on the results of the previous section, it is possible to trace the origin of each solution. Further we will indicate the predominant contributions (IS, IV, SS, SV, SF).

\subsubsection{$K^\pi=0^-$}\label{M2K0}

The characteristic equation of the set (\ref{mu=0(E)}) describing $K^\pi=0^-$ {\bf electric} excitations has four eigenvalues.
The $E10$ strength is calculated for each state.
\begin{table}[ht!]
\caption{The electric $K^{\pi}=0^-$ excitations of $^{164}$Dy.}
\begin{ruledtabular}\begin{tabular}{cccc}
  & &    $E_{10}$  (MeV)     & $B(E10)$ ($e^2$fm$^2$)     \\
 \cline{3-4}
 & &  \multicolumn{2}{c}{$\xi_0^{K=1}/\tilde\xi_0^{K=1}=1$}\\
 IS & SS &    ~0.00  &  --        \\
 IS & SF &    ~7.69  & ~0.05      \\
 IV & SF &    11.64  & ~3.65      \\
 IV & SS &    13.19  & 11.17      \\[1mm]
\hline
  & &  \multicolumn{2}{c}{$\xi_0^{K=1}/\tilde\xi_0^{K=1}=-30$}\\
 IS & SS &    40.61  & ~0.00     \\
 IS & SF &    ~7.57  & ~0.00     \\
 IV & SF &    11.57  & ~4.93     \\
 IV & SS &    13.01  & 10.73     \\
\end{tabular}\end{ruledtabular}
\label{tab4}
\end{table}
Results of calculations are shown in the Table~\ref{tab4}.
As before, two methods are used to eliminate the excitation of the center of mass motion.
The 13.19 (13.01) MeV state is the $K^\pi=0^-$ branch of GDR determined by the dynamics of spin-scalar variables $\bar\R_0^+$, $\bar\P_0^+$.
The way that moves the CMM energies to the high energy region is more preferable, since it allows one to completely avoid the admixture of spurious (IS SS) $0^-$ state (see the bottom part of the Table~\ref{tab4}).
In this case, all excitations in sum exhaust 100\% of the dipole sum rule~(\ref{SRE1}).
So from now on we will follow to this way.

Excitations at energies 7.57 MeV and 11.57 MeV are generated by the variables
$\R_0^E$ and $\P_0^E$ defined by eq.~(\ref{RP0E}).
It can be shown that $\R_0^E$ is the average value ​​of the spin-dipole operator $s_{10}$ (\ref{sd}):
\begin{eqnarray}
\langle s_{10}\rangle&=&-\frac{i}{2\sqrt2}{\rm Tr}\left(\hat\rho[\hat \br\times\hat\sigma]_z\right)=
-\frac{i}{2\sqrt2}\langle \hat x\hat\sigma_y-\hat y\hat\sigma_x\rangle
\nonumber\\
&=&-\frac{1}{2}\left(R^\u_1+R^\d_{-1}\right)=
-\frac{1}{2}R_0^E,\label{R0E}
\end{eqnarray}
where $\hat\rho$ is a density matrix and $\hat\sigma_i$ are Pauli matrices~\cite{Var}. Analogously,
\begin{eqnarray}
&&-\frac{i}{2\sqrt2}{\rm Tr}\left(\hat\rho[\hat \bp\times\hat\sigma]_z\right)
=
-\frac{i}{2\sqrt2}\langle \hat p_x\hat\sigma_y-\hat p_y\hat\sigma_x\rangle
\nonumber\\
&&=-\frac{1}{2}\left(P^\u_1+P^\d_{-1}\right)=
-\frac{1}{2}P_0^E.\label{P0E}\quad
\end{eqnarray}
Expressions (\ref{R0E}), (\ref{P0E}) show that electric spin-dipole excitations
(oscillations along the $z$-axis)
are caused by spin-flip transitions.
The procedure used to eliminate the contribution of the CMM also
depresses  the isoscalar spin dipole state (compare IS SF lines in the  Table~\ref{tab4}).
The state with energy 11.57 MeV is the isovector Electric Spin Dipole Resonance (ESDR).

The energy centroid of  two isovector electric excitations (GDR and ESDR), $\bar E_{10}=12.56$ MeV, and the sum of their excitation probabilities $\sum B(E10)=15.66\ e^2$fm$^2$
are almost identical to those obtained for GDR in the $\eta=0$ approximation (compare with the Table~\ref{tab2}). ESDR is not excited when $\eta=0$ (see section~\ref{K=0,eta=0}).
The inclusion of the spin-orbit potential results in a slight change in the energy of both modes. The energy of GDR changes from 12.56 MeV ($\eta=0$) to 13.01 MeV. At the same time, the energy of ESDR changes from 12.02 MeV  to 11.57 MeV.
Herewith, part of the $E1$ strength of GDR is transferred to the ESDR, see Table~\ref{tab4}.
So, the inclusion of the spin-orbit potential opens the channel by which the ESDR takes away
the essential part of the strength from the GDR.
The gap between the GDR and the ESDR is $\Delta_{\rm sd}=E_{\rm GDR}-E_{\rm ESDR}=0.54$ MeV when $\eta=0$, and  it increases to $\Delta_{\rm sd}=1.44$ MeV in exact calculations.

\begin{figure}[t!]
\includegraphics[width=\columnwidth]{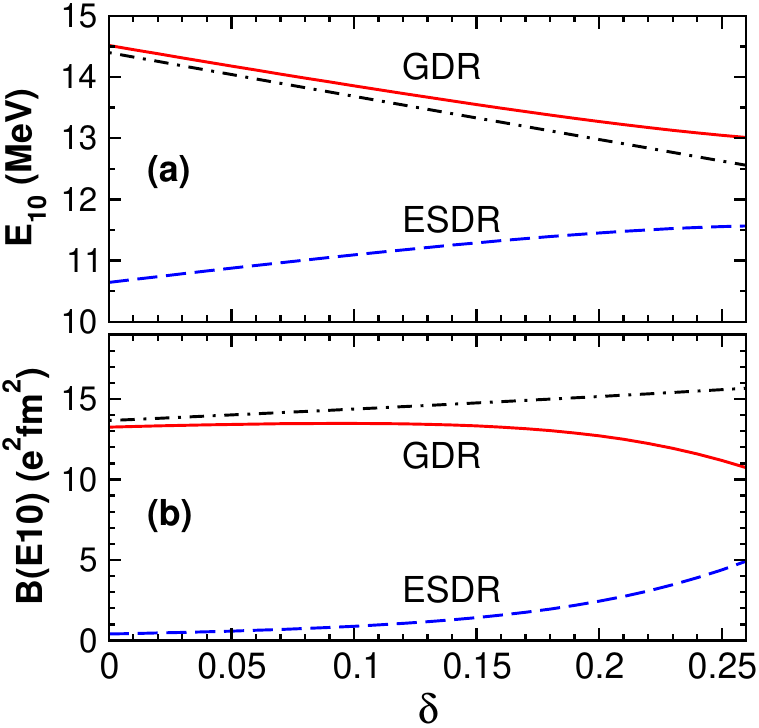}
\caption{Energies (a) and $E1$ strengths (b) of $K^{\pi}=0^-$ GDR (red line) and ESDR (blue dashed line)
vs. deformation $\delta$.
The energy centroid $\bar E_{10}$ and summed $B(E10)$ value are shown by black dot-dashed lines. Calculations are performed for $^{164}$Dy.}
\label{fig4}
\end{figure}
The dependence of the GDR and ESDR characteristics on the deformation is shown in the Figure~\ref{fig4}.
As can be seen, with the deformation increasing the splitting $\Delta_{\rm sd}$ decreases, while  the share of the strength attributed to ESDR increases, its
contribution to EWSR increasing from 2.21\% for $\delta=0$ to 28.98\% for $\delta=0.26$.
In the spherical limit, the splitting is maximal, $\Delta_{\rm sd}=3.87$ MeV, while the excitation probability is only $1.23\  (=3\cdot 0.41)\ \e^2$fm$^2$.
The energy centroid shifts toward lower energies from 14.40 MeV when $\delta=0$ to 12.56 MeV for $\delta=0.26$, the summed $E10$ strength increases  from 13.66 $e^2$fm$^2$ to 15.66 $e^2$fm$^2$.

\begin{table}[h!]
\caption{The magnetic
$K^\pi=0^-$ excitations of $^{164}$Dy.}
\begin{ruledtabular}\begin{tabular}{cccc}
  &   & $E_{20}$  (MeV)  & $B(M20)$ ($\mu_{N}^2$fm$^2$) \\
\hline
  IS & \multirow{2}{*}{SV} &~4.79 & ~141.75 \\
  IV &  & 10.61 & 1277.45 \\
\hline
 IS & \multirow{2}{*}{SF} &~7.61 & ~~4.96 \\
 IV &  & 12.02 & 588.97 \\
\end{tabular}\end{ruledtabular}
\label{tab5}
\end{table}

The characteristic equations of the set (\ref{mu=0(M-SV)}) and the set (\ref{mu=0(M-SF)}) describe the SV and SF $K^\pi=0^-$ spin {\bf magnetic} quadrupole responses, respectively.
Consequently, the variables $\R_0^-, P_0^-$ (the set \ref{mu=0(M-SV)})) generate SV excitations, whereas the variables $\R_0^M, P_0^M$ (the set (\ref{mu=0(M-SF)})) generate SF excitations.
Results of calculations are shown in the Table~\ref{tab5}.
It can be shown that $\R_0^-$ and $\R_0^M$ are the simple combinations of
average values ​​of spin-dipole operators $s_{20}$ and $s_{00}$ (\ref{sd}):

\begin{eqnarray}&&\langle\hat s_{20}\rangle=\frac{1}{\sqrt6}\left[\R_0^-+\frac{\R_0^M}{\sqrt2}\right],\\
&&\langle\hat s_{00}\rangle=-\frac{1}{\sqrt6}\left[\frac{\R_0^-}{\sqrt2}-\R_0^M\right].
\end{eqnarray}
 The summed $\sum B(M20)$ strength over all four states  amounts 2013.13 $\mu_{N}^2$fm$^2$. The magnetic EWSR (\ref{SR_M20}) is completely exhausted, with 63.5\%  coming from a strong peak at 10.61 MeV.

\subsubsection{$K^\pi=1^-$}\label{Kpi1}

In deformed nuclei the electric and magnetic channels of $1^-$ excitation are coupled.
That is why the dynamical equations for SS variables
$\bar\R^+_1$,$\bar\P^+_1$, $\R^+_1$, $\P^+_1$ and
SV ($\bar\R^-_1$,$\bar\P^-_1$, $\R^-_1$, $\P^-_1$)
and SF ($\bar\R^\d_0$,$\bar\P^\d_0$, $\R^\d_0$, $\P^\d_0$)
variables are also coupled (see eq.~(\ref{mu=1(12)})).
The characteristic equation of this system has six eigenvalues.
They are shown in the Table~\ref{tab6} together with
the $B(E11)$ and $B(M21)$ reduced transition probabilities.

\begin{table}[ht!]
\caption{The results of WFM calculations for $^{164}$Dy: energies $E_{\lambda1}$ of $1^-$ excitations
and $B(E11)\!\!\uparrow$ and $B(M21)\!\!\uparrow$ transition probabilities.
The spin-vector contributions to $B(M21)\!\!\uparrow$  ($\sim\bar\R^-_1,\ \R^-_1$) are indicated in brackets.}
\begin{ruledtabular}\begin{tabular}{ccccc}
  & &    $E_{\lambda1}$     & $B(E11)\!\!\uparrow$   & $B(M21)\!\!\uparrow$    \\
  & &   (MeV)   & ($e^2$fm$^2$)    & ($\mu_{N}^2$fm$^2$)   \\
 \cline{3-5}
  & &  \multicolumn{3}{c}{$\xi_0^{K=1}/\tilde\xi_0^{K=1}=1$}\\
 IS & SS &   ~0.00  &  --    & ~~~~0.00              \\
 IS & SF &    ~4.86  & ~0.03  & ~214.42  (~~~0.01)    \\
 IS & SV &    ~7.65  & ~0.01  & ~~12.51  (~~14.61)    \\
 IV & SF &    10.57  & ~0.17  & 1793.50  (~~~~1.54)   \\
 IV & SV &    11.97  & ~0.26  & 1887.98  (1745.60)    \\
 IV & SS &    16.37  & 23.42  & ~~~~0.92              \\[1mm]
\hline
  & &  \multicolumn{3}{c}{$\xi_0^{K=1}/\tilde\xi_0^{K=1}=-30$}\\
 IS & SS &    52.44  & ~0.00  & ~~~~0.00              \\
 IS & SF &    ~4.77  & ~0.00  & ~212.19  (~~~0.00)    \\
 IS & SV &     ~7.59  & ~0.00  & ~~15.77  (~~15.01)    \\
 IV & SF &      10.57  & ~0.17  & 1794.12  (~~~~1.53)   \\
 IV & SV &    11.97  & ~0.28  & 1887.98  (1744.10)    \\
 IV & SS &    16.25  & 23.90  & ~~~~1.08              \\
\end{tabular}\end{ruledtabular}
\label{tab6}
\end{table}

The inclusion of the spin-orbit potential leads to the negligible increase in the GDR energy from 16.28 MeV to 16.37 MeV (or from 16.16 MeV to 16.25 MeV)
and to the decrease of the $E1$ strength about 1$\%$ (compare the results in Table~\ref{tab6} with those in Table~\ref{tab1}).
That is, the GDR is weakly sensitive to a spin-orbit potential.

Zeroing the CMM excitation energy yields an exhaustion of the dipole EWSR~(\ref{SRE1}) by 98.73\%. The rest 1.27\% are taken by the CMM.
The calculations exhaust 100\% of the EWSR
when the energy of the CMM is pushed to the high energy region.
This way wins again  because it avoids the admixture of a spurious state (see IS SS in the bottom part of the Table~\ref{tab6}). The largest contribution to the EWSR,
98.68\%, comes from the GDR.

The summed $M21$ strength amounts 3911.14 $\mu_{N}^2$fm$^2$ and the magnetic EWSR (\ref{SR_M20}) is exhausted by 100\%. The share of GDR is less than 1\%.

 The detailed analysis of the structure of magnetic $1^-$ excitations revealed that the SV contribution dominates in the 7.59 MeV and 11.97 MeV transitions, while the SF contribution dominates in the 4.77 MeV and 10.57 MeV transitions (see Table~\ref{tab6} where SV contribution to $B(M21)$ is given in brackets).
 The influence of the spin-orbit potential can be evaluated by comparing the results shown in Tables~\ref{tab6} with the results of the  Table~\ref{tab3}.
 As can be seen, the excitation energies remain almost the same.
 A noticeable result is that  the $M21$ strength of isovector states is redistributed in favor of the SV ones.

 It was shown in the previous section that the
 ESDR is  mediated by the operator $s_{10}$, which generates spin-flip transitions.
For the operator $s_{11}$ the following holds:
\begin{eqnarray}
\langle s_{11}\rangle=-\frac{1}{\sqrt2}\left(R^-_1+\sqrt2R^\d_0\right).\label{R1E}
\end{eqnarray}
Unlike $\langle s_{10}\rangle$, in addition to the spin-flip, there is also the spin-vector
contribution.
Herewith, the magnetic response is determined by
\begin{eqnarray}
\langle s_{21}\rangle=\frac{1}{2\sqrt2}\left(R^-_1-\sqrt2R^\d_0\right).\label{R2K}
\end{eqnarray}
Thus, in contrast with $0^-$, the same variables govern both the electrical and magnetic response.
Two $K^{\pi}=1^-$ isovector states with energies 10.57 MeV and 11.97 MeV exhaust 97 \% of the magnetic EWSR.
They also take 1.3 \% of the electrical EWSR.
Both the spin-flip and the spin-vector oscillations contribute to the formation of these states.
The analysis revealed that the spin-flip response ($\sim \bar R^\d_0$) dominates at 10.57 MeV excitation.
The spin-vector vibrations ($\sim \bar R^-_1$)
produce the dominant contribution to the 11.97 MeV excitation, see Table~\ref{tab6}.
In general, the electric spin response is suppressed by the magnetic one in electromagnetic transitions.

\subsubsection{$K^\pi=2^-$}
The system of equations (\ref{mu=2(4)}) describing the $K^\pi=2^-$ component of the spin $M2$ resonance coincides with eqs.~(\ref{K=1sv(eta=0)}),
the difference being again only in the notations of variables:
\begin{eqnarray}
\dot {\bar\R}_1^\d&=&
\frac{1}{m}\bar\P_1^\d,
\nonumber \\
\dot {\bar\P}_1^\d&=&
-m\omega^2C_1\bar\R_1^\d  - \frac{2}{3}\alpha\delta Xm\tilde\omega^2\R_1^\d
\nonumber\\&&
-\frac{A}{4}\left(\chi_1 \bar\R_1^\d + \chi_0 X\R_1^\d\right)
\nonumber \\
\dot {\R}_1^\d&=&
\frac{1}{m}\P_1^\d,
\nonumber \\
\dot {\P}_1^\d&=&
-m\omega^2C_1\R_1^\d   - \frac{2}{3}\alpha\delta Xm\tilde\omega^2\bar\R_1^\d
\nonumber\\&&
-\frac{A}{4}\left(\chi_0 \R_1^\d + \chi_1 X\bar\R_1^\d\right).
\label{M22(etano)}\end{eqnarray}
The energies of $2^-$ excitations and respective  transition probabilities
are listed in the Table~\ref{tab7}. The summed strength amounts 3563.60 $\mu_{N}^2$fm$^2$.
The magnetic EWSR (\ref{SR_M20}) is exhausted by 100\%.
\begin{table}[h]
\caption{The magnetic $K^\pi=2^-$ excitations of $^{164}$Dy.}
\begin{ruledtabular}\begin{tabular}{cccc}
& & $E_{22}$  (MeV)  & $B(M22)\!\!\uparrow$ ($\mu_{N}^2$fm$^2$) \\
\cline{3-4}
 IS & SF &  ~7.61 & ~~29.77 \\
 IV & SF &   12.02 & 3533.83 \\
\end{tabular}\end{ruledtabular}
\label{tab7}
\end{table}

One can note that energies of $K^\pi=2^-$ SF excitations (Table ~\ref{tab7}) coincide with that of
$K^\pi=0^-$ SF excitations (Table~\ref{tab5}). It looks not surprising, because the equations describing both excitations coincide, that happens in its turn due to the neglect by the dynamics of
third rank moments. As we know, taking them into account leads to the appearance of seven $1^-$ excitations of different nature, including GDR (see Introduction and Figs.~\ref{fig1} and~\ref{fig2}).
It can be expected that in this case too the above-mentioned degeneration will be removed.

\section{Summary and discussion}\label{IV}

We have found in $^{164}$Dy nucleus 14
levels located in the excitation energy range of $4–17$ MeV. Of these, seven are $K^\pi=0^-$ excitations, five are $K^\pi=1^-$ and two are $K^\pi=2^-$.

The highest energy states are the two branches of the GDR.
   This is the isovector spin-scalar
mode. Its isoscalar counterpart was artificially put to the zero (or very large) energy to
exclude the center of mass motion, which represents the spurious mode.
   The energies of both GDR branches were described
by fitting the parameters of the dipole-dipole interaction. The $K^\pi=0^-$ and $K^\pi=1^-$ peak positions were found at
13.01~MeV and 16.25~MeV, respectively. The deformation-induced splitting is 3.24~MeV. These results are aligned with the established patterns observed in the physics of the GDR~\cite{BMII,Atlas_GDR}.
\begin{figure}[t!]
\includegraphics[width=\columnwidth]{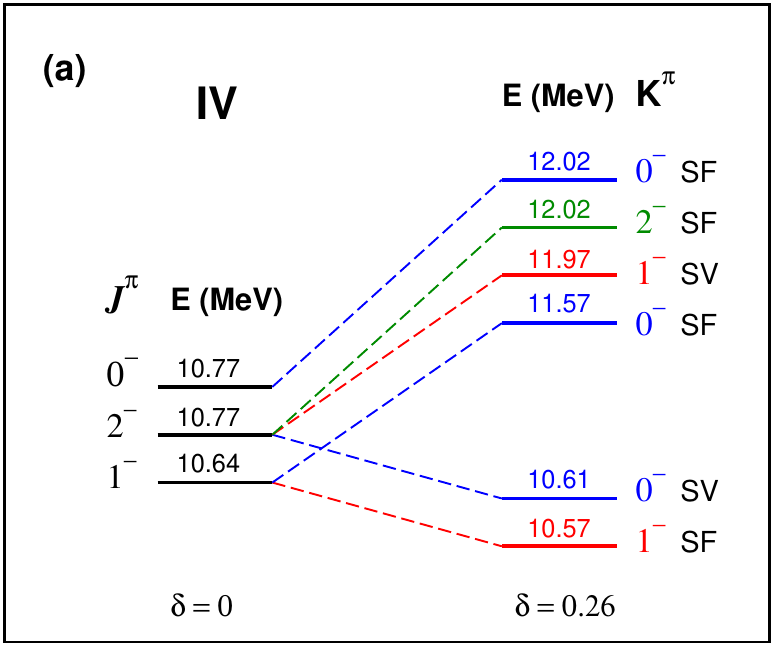}\\\includegraphics[width=\columnwidth]{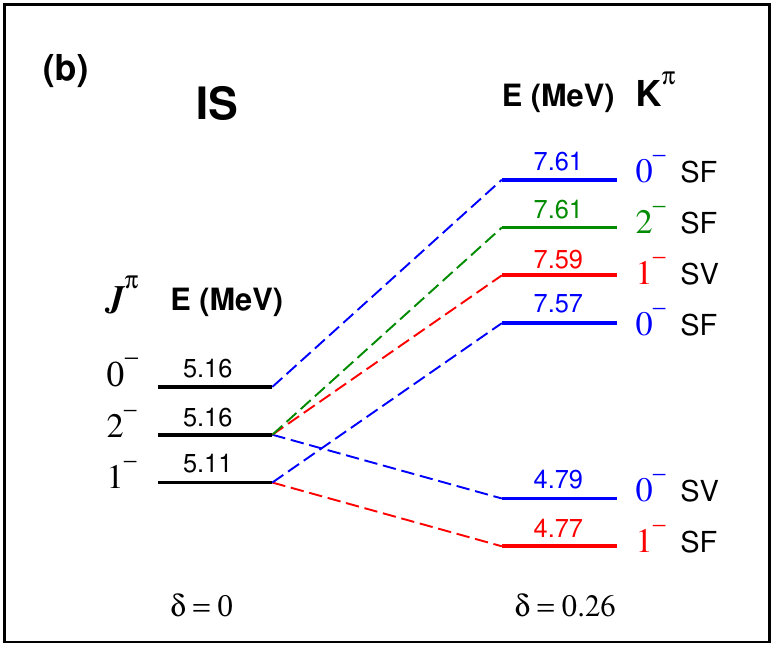}
\caption{Energies of IV (a) and IS (b) spin-dipole excitations for $\delta=0$ (see Table~\ref{tab8}) and for $\delta=0.26$
(see Tables~\ref{tab4},~\ref{tab5},~\ref{tab6} and~\ref{tab7}). Calculations are performed for $^{164}$Dy.}
\label{fig5}
\end{figure}
\begin{table}[b!]
\caption{The excitations of a hypothetical spherical nucleus with N and Z corresponding to $^{164}$Dy.}
\begin{ruledtabular}\begin{tabular}{ccccc}
 & $J^\pi$ & $E(J^\pi)$  & $B(E1)\!\!\downarrow$ & $B(M2)\!\!\downarrow$  \\
 &  & (MeV) & ($e^2$fm$^2$) & ($\mu_{N}^2$fm$^2$) \\
\cline{2-5}
    & $0^-$ & 10.77 & 0    & 0 \\
 IV & $1^-$ & 10.64 & 0.41 & 0 \\
    & $2^-$ & 10.77 & 0    & 1954.06 \\
\cline{2-5}
    & $0^-$ & ~5.16 & 0    & 0 \\
 IS & $1^-$ & ~5.11 & 0    & 0 \\
    & $2^-$ & ~5.16 & 0   & ~62.27 \\
\end{tabular}\end{ruledtabular}
\label{tab8}
\end{table}
Spin resonances induced by the spin-dipole operator~(\ref{sd}) lie lower in energy.
   They are grouped in four multiplets, see Fig.~\ref{fig5}.

   The excitations with
$K^{\pi}=2^-$ (spin index SF, energy $E_{22}=12.02$ MeV,
Table~\ref{tab7}),
$K^{\pi}=1^-$ (spin index SV, energy $E_{21}=11.97$ MeV,
Table~\ref{tab6}) and
$K^{\pi}=0^-$ (spin index SV, energy $E_{20}=10.61$ MeV,
Table~\ref{tab5}) form the isovector
triplet. In the spherical limit these levels converge in one point with quantum numbers
$J^{\pi}=2^-$ and energy $E_2=10.77$ MeV (see Table~\ref{tab8} and Fig.~\ref{fig5}(a)).
The levels $K^{\pi}=2^-$ and $K^{\pi}=0^-$ are pure magnetic ones.
Their excitation probabilities are
$B(M22)\!\!\uparrow=3533.83$ $\mu_N^2$fm$^2$ (see Table~\ref{tab7}) and
$B(M20)=1277.45$ $\mu_N^2$fm$^2$ (see Table~\ref{tab5}).
The level $K^{\pi}=1^-$ has the predominantly magnetic nature with the small electric
admixture. The respective excitation probabilities are
$B(M21)\!\!\uparrow=1887.98$ $\mu_N^2$fm$^2$ and
$B(E11)\!\!\uparrow=0.28$ $e^2$fm$^2$ (see Table~\ref{tab6}).

The isoscalar counterpart of this triplet is formed by the excitations with energies
$E_{22}=7.61$ MeV ($K^{\pi}=2^-$, SF, Table~\ref{tab7}),
$E_{21}=7.59$ MeV ($K^{\pi}=1^-$, SV, Table~\ref{tab6}) and
$E_{20}=4.79$ MeV ($K^{\pi}=0^-$, SV, Table~\ref{tab5}).
In the spherical limit these levels converge in one $J^{\pi}=2^-$ level with the energy
$E_2=5.16$ MeV (Table~\ref{tab8} and Fig.~\ref{fig5}(b)).
The levels $K^{\pi}=2^-$ and $K^{\pi}=0^-$ are pure magnetic ones.
Their excitation probabilities $B(M22)\!\!\uparrow=29.77$ $\mu_N^2$fm$^2$ and
$B(M20)=141.75$ $\mu_N^2$fm$^2$ are the order of magnitude less than in the isovector
case. The level $K^{\pi}=1^-$ also has the pure magnetic nature with
$B(M21)\!\!\uparrow=15.77$ $\mu_N^2$fm$^2$. Contrary to its isovector counterpart it has not any
electric admixture. However the last remark depends on the method of excluding the CM motion.
As it is seen from the Table~\ref{tab6}, in the case of zeroing CMM excitation energy this level has
a little bit another energy $E_{21}=7.65$ MeV and
the nonzero (though rather small) $B(E11)$ value.

Further, the excitations with energies
$E_{10}=11.57$ MeV ($K^{\pi}=0^-$, SF, Table~\ref{tab4}) and
$E_{11}=10.57$ MeV ($K^{\pi}=1^-$, SF, Table~\ref{tab6}) form the isovector doublet.
In the spherical limit these levels converge in one $J^{\pi}=1^-$ level with the energy
$E_1=10.64$ MeV, see Fig.~\ref{fig5}(a).
The level $K^{\pi}=0^-$ is pure electric one (ESDR). Its excitation probability is
$B(E10)=4.93$ $e^2$fm$^2$ (see Table~\ref{tab4}). The level $K^{\pi}=1^-$ has the
predominantly magnetic nature with the very small electric admixture.
The respective excitation probabilities are
$B(M21)\!\!\uparrow=1794.12$ $\mu_N^2$fm$^2$ and
$B(E11)\!\!\uparrow=0.17$~$e^2$fm$^2$ (Table~\ref{tab6}).

The isoscalar counterpart of this doublet is formed by the excitations with energies
$E_{10}=7.57$ MeV ($K^{\pi}=0^-$, SF, Table~\ref{tab4}) and
$E_{11}=4.77$ MeV ($K^{\pi}=1^-$, SF, Table~\ref{tab6}).
In the spherical limit they converge in one $J^{\pi}=1^-$ level with the energy
$E_1=5.11$ MeV (Table~\ref{tab8} and Fig.~\ref{fig5}(b)). The conclusion about the electromagnetic nature of these states depends on the method of excluding the CM motion. As it is seen from the Table~\ref{tab4} the level $E_{10}=7.57$ MeV
is not excited electromagnetically in the case of the large CMM energy $E_{\rm{CM}}$, whereas in the
case of $E_{\rm{CM}}=0$ it has already a little bit another energy $E_{10}=7.69$ MeV and
the nonzero (though rather small) $B(E10)$ value. According to the
Table~\ref{tab6} the level $E_{11}=4.77$ MeV is pure magnetic one if $E_{\rm{CM}}$ is large, whereas it
acquires the electric property (though rather weak one) and the slightly another energy
$E_{11}=4.86$ MeV, if $E_{\rm{CM}}=0$.

\begin{figure}[t!]
\includegraphics[width=\columnwidth]{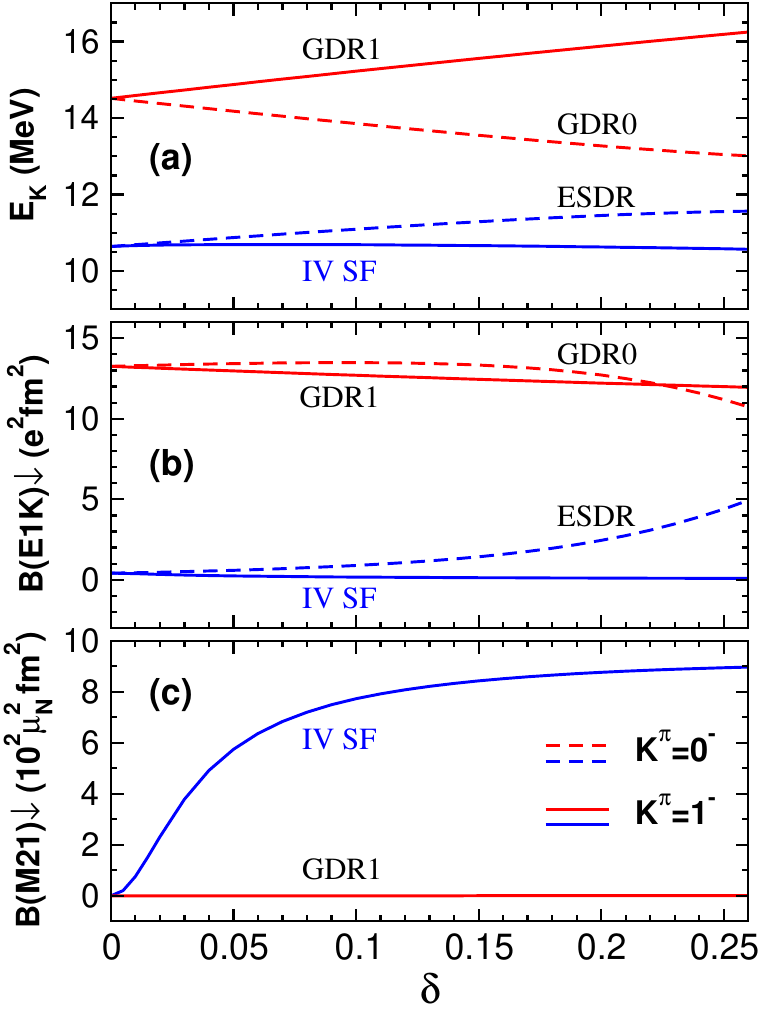}
\caption{(a) Energies of $K^\pi=0^-$ GDR (GDR0, red dashed line),
 $K^\pi=1^-$ GDR (GDR1, red line), $K^\pi=0^-$ ESDR (blue dashed line) and $K^\pi=1^-$ state (IV SF, blue  line) with
(b) $B(E1K)$ and (c) $B(M21)$ strengths
vs.~deformation~$\delta$. Calculations are performed for $^{164}$Dy.}
\label{fig6}
\end{figure}
In the case of $K^\pi=0^-$, the electric and magnetic modes are separated.
For $K^\pi=1^-$ excitations the situation is different.
In deformed nuclei, the electric and magnetic channels of $1^-$ excitation are coupled. The result is non-zero $B(E11)$ and
$B(M21)$ transition probabilities for each $1^-$ level,
see Table~\ref{tab6}.
In the spherical limit there is no mixing between the $J^\pi=1^-$ and $J^\pi=2^-$.
The interplay between GDR, ESDR (which is IV SF $0^-$) and IV SF $1^-$ state
is shown in the Fig.~\ref{fig6}.
It is noteworthy that the energy and $E11$ strength of the IV SF
 $1^-$ state
depend weakly on the deformation, whereas the $B(M21)$, on the contrary, drops sharply, reaching zero at $\delta = 0$.
It is necessary also to note the remarkable behaviour of this level with deformation: beginning at
the zero deformation as the pure electric 
excitation with the $B(E1)\!\!\uparrow=1.23\ e^2$fm$^2$
it ends at the deformation $\delta=0.26$ as the practically pure magnetic state with
the $B(E11)\!\!\uparrow=0.17 e^2$fm$^2$ and the $B(M21)\!\!\uparrow=1794\ \mu_N^2$fm$^2$.


The $M2$ transition strength is dominated by the isovector response.
The isoscalar one has a small contribution. It mainly makes visible the transitions in the energy range of $4-8$ MeV.
In the energy range above 8 MeV,
the spectrum of isovector $1^-$  magnetic quadrupole excitations exhibits two main peaks at 10.57 MeV (SF) and 11.97 MeV (SV) with nearly equal $B(M21)$ strengths (see Table~\ref{tab6}).
Their energy centroid is at 11.29 MeV.
The spectrum of isovector $0^-$ magnetic excitations consists of two levels at 10.61 MeV (SV) and 12.02 MeV (SF), see Table V.
Their energy centroid is at 11.06 MeV.
The splitting with the $1^-$ bump is only 0.23 MeV.
The $2^-$ component is represented by one strong SF state at 12.02 MeV (see Table~\ref{tab7}).

To sum it up, the $K^\pi=0^-,\ 1^-$ and $2^-$ centroids of the isovector spin $M2$ strength are found at energies 11.06 MeV, 11.29 MeV and 12.02 MeV, respectively.
So, the deformation-induced splitting of the $M2$ giant resonance is noticeably smaller than that of the $E1$ giant resonance.
This conclusion is consistent with the results of the paper~\cite{Nester-M2}, where the distribution of the $M2$ strength in heavy deformed nuclei was studied within the RPA framework.

Theoretically, ESDR and GDR should be observed together since they have the same $J^\pi=1^-$ value
and are located in the same excitation energy region.
Spherical atomic nuclei are preferred candidates for searching for ESDR, 
since in this case the most pronounced gap with GDR is expected and there is no mixing with $2^-$ states.
It is interesting to find experimental indications to support our prediction.
The collective spin-dipole excitations were the subject of interest in the paper~\cite{Morsch78}, where $^{208}$Pb($p,p^\prime$) reaction was studied. The two states at 6.26 and 8.37 MeV were interpreted as $1^-$ spin-flip excitations.
In a recent paper~\cite{Pb_GDR+}, photoneutron reactions on $^{208}$Pb in the GDR energy region has been investigated. Several resonance structures have been detected in low-energy $^{208}$Pb($\gamma$, $n$) cross sections. Thus, in particular, the broad shoulder at 9.45 MeV and a prominent peak at $\sim10$ MeV have been observed. The ESDR position in $^{208}$Pb calculated within the WFM method is 9.83 MeV~\cite{BaMoIJMPE26} and can be assigned to one of the above states.
The $^{208}$Pb($p,p^\prime\gamma$) reaction has been studied in Ref.~\cite{Krakow_GDR+}. Several discrete states of electric nature have been detected between 7 and 10 MeV. It was supposed that some of them are collective $1^-$ spin-flip excitations. That is certainly in line with our findings.

It is worth to note on the remarkable analogy between
 $1^-$ excitations and $1^+$ scissors modes, see Ref.~\cite{BaMoPRC22}. In both cases there are four types of
collective motions which are classified as isoscalar  spin-scalar, isovector  spin-scalar,
isoscalar  spin-vector and isovector  spin-vector.

The rotational isoscalar spin-scalar motion is the rotation of a nucleus as a whole and
represents the conservation of the nucleus angular momentum. The translational
isoscalar  spin-scalar motion is the linear motion of a nucleus as a whole and represents
the conservation of the nucleus linear momentum.

Furthermore, the rotational isovector  spin-scalar motion is the conventional scissors mode.
This excitation has the appreciable admixture of the non-rotational motion (quadrupole dynamical deformations), due to which it has both magnetic and electric nonzero excitation
probabilities, $B(M1)$ and $B(E2)$ respectively. The translational
isovector  spin-scalar motion is the Giant Dipole Resonance.
This excitation has not any
rotational admixture. Nevertheless, due to the spin-orbit potential it has (additionally to the electrical excitation probability $B(E1)$) the small 
magnetic excitation
probability $B(M2)$, see Table~\ref{tab6}.

The rotational isoscalar  spin-vector motion is one of two spin scissors. It has the largest
(among three scissors) magnetic dipole excitation probability $B(M1)$ and the smallest (order
of magnitude) electric excitation probability $B(E2)$, see Tabe I in Ref. \cite{BaMoPRC22}. One can
say that this scissors represents practically pure rotational motion of spin-up nucleons with
respect of spin-down nucleons, the admixture of the non-rotational one being negligibly small.
This means that it is practically pure magnetic mode. The translational
isoscalar spin-vector motion is identified as one of $1^-$ excitations with the energy
$E=7.59$ MeV (see Table~\ref{tab6}). It represents the linear oscillations of all spin-up nucleons with
respect of all spin-down nucleons. It is interesting to note that having zero $B(E11)$ and
nonzero $B(M21)$ this mode also turns out pure magnetic one.

The rotational isovector spin-vector motion is the second of two spin scissors (complicate
spin scissors) with a little bit smaller $B(M1)$. Its $B(E2)$ value is one order of magnitude
bigger than that of the first (isoscalar  spin-vector) spin scissors. This fact says that
this mode has the appreciable admixture of the non-rotational motion, exactly as the
conventional scissors.
The translational isovector spin-vector motion is identified as one more $1^-$ excitation
with the energy  $E=11.97 $ MeV (Table~\ref{tab6}). It has rather big magnetic quadrupole excitation
probability $B(M21)=1887$ $\mu_N^2fm^2$. Analogously to the complicate scissors it can be
described as the out of phase linear oscillations of spin up protons + spin down neutrons with respect of spin down protons + spin up neutrons. However, contrary to the scissors this mode
has very small  $B(E11)$, so one can count it as the pure magnetic one.
The spin-vector motion contribution disappear in the spherical limit (like spin scissors).

Now one can say that our expectations, announced  in the Introduction, concerning the possible existence of spin giant resonances of various multipolarities is justified. We have found the
sought out isovector and isoscalar spin-vector giant dipole resonances.

Here is the proper place to clarify the relation between the spin-flip (SF) and spin-vector
(SV) types of excitations. There exists the popular opinion that SF and SV are just the
different names of the same phenomenon. See, for example, the paper \cite{Morsch78}: "... a collective dipole spin-flip excitation ($L=1,\  S=1$) which in a simple collective picture can be interpreted as a collective dipole oscillation of spin-up particles against
spin-down particles." However, as we have shown, SF and SV represent two different types of excitations.
Moreover, the transition strengths are 
quite clearly
separated into the spin-ﬂip and spin-vector components.

\section{Conclusions}\label{V}

In this work we have investigated the properties of the GDR and electric and magnetic spin-dipole resonances in heavy even-even axial deformed nuclei within the framework of the WFM method.

The energy position and caused by the deformation splitting of GDR are calculated for $^{164}$Dy. The obtained values ​​are in good agreement with the general global trends for the GDR.

Our calculations show that
spin resonance can be induced by an alternating electric field.
This mode (ESDR) is low-energy satellite of the GDR and is not excited if the spin-orbit potential is neglected.
We have studied experimental $1^-$ spectra of spherical nuclei in search of states that could correspond to the ESDR. Several suitable candidates have been found.

Of considerable interest are spin-vector states, interpreted as collective dipole translational motion of spin “up” nucleons relative to spin “down” nucleons. These oscillations are the translational analogue of the $1^+$ spin scissors mode.
It was found that spin-vector displacements are generated by the spin part of the $\hat O(M2,1)$ operator.

We have evaluated the energy, transition probability and
deformation-induced splitting of the spin $M2$ giant resonance.
The third-rank tensors are required to calculate the orbital responce.
However, the presented results are seems vital since there are calculations indicating that the magnetic quadrupole strength is dominated by the spin response~\cite{Paar}.

\begin{acknowledgments}
Valuable discussions with V.~Nesterenko and R.~Nazmitdinov are gratefully acknowledged.
\end{acknowledgments}

\begin{widetext}

\appendix
 \section{Mean field}
\label{MeanF}

 The contribution of $H_{qq}$ (\ref{Hqq}) to the mean field potential is easily
found by replacing one of the $q_{2\mu}$ operators by the average value.
We have \cite{BaSc}
\begin{equation}
\label{potenirr}
V^{\tau+}_{\rm qq}=\sqrt6\sum_{\mu}(-1)^{\mu}Z_{2-\mu}^{\tau +}q_{2\mu},
\end{equation}
where $\tau$ denotes the isospin index ($\tau={\rm n, p}$) and
\begin{eqnarray}
\label{Z2mu}
Z_{2\mu}^{{\rm n}+}=\kappa R_{2\mu}^{{\rm n}+}
+\bar{\kappa}R_{2\mu}^{{\rm p}+},\quad
Z_{2\mu}^{{\rm p}+}=\kappa R_{2\mu}^{{\rm p}+}
+\bar{\kappa}R_{2\mu}^{{\rm n}+}.
\end{eqnarray}

The analogous expression for $H_{\rm dd}$ (\ref{Hdd}) is found in a
standard way \cite{BaMoPRC13} with the following result for
the Wigner transform of the mean field:
\begin{equation}
V^{\rm n+}_{\rm dd}=\xi \V^{\rm n+}_{\rm dd}+\bar\xi \V^{\rm p+}_{\rm dd},\quad
V^{\rm p+}_{\rm dd}=\xi \V^{\rm p+}_{\rm dd}+\bar\xi \V^{\rm n+}_{\rm dd},
\end{equation}
with
\begin{eqnarray}
\V^{\tau+}_{\rm dd}=\sum_{\mu=-1}^{1}(-1)^{\mu}R^{\tau+}_{-\mu}r_\mu,
\quad
\V^{\tau-}_{\rm dd}=\V^{\tau\d}_{\rm dd}=\V^{\tau\u}_{\rm dd}=0.
\end{eqnarray}

For the spin-dipole mean-ﬁeld potential, we derived:
\begin{equation}
V^{\tau+}_{\rm sd}=0,\quad
V^{\rm n\zeta}_{\rm sd}=\chi \V^{\rm n\zeta}_{\rm sd}+\bar\chi \V^{\rm p\zeta}_{\rm sd},\quad
V^{\rm p\zeta}_{\rm sd}=\chi \V^{\rm p\zeta}_{\rm sd}+\bar\chi \V^{\rm n\zeta}_{\rm sd},
\end{equation}
where $\zeta\!=-,\,\uparrow\downarrow,\,\downarrow\uparrow$ and
\begin{eqnarray}\label{Vsd}
&&\V^{\tau-}_{\rm sd}=
\frac{1}{3}r_0\left(\M_{2,0}^\tau+\frac{1}{2}\M_{0,0}^\tau\right)
-\frac{1}{4}\left(r_{-1}\M_{2,1}^\tau+r_1\M_{2,-1}^\tau\right)
-\frac{1}{4}\left(r_{-1}\E_{1,1}^\tau+r_1\E_{1,-1}^\tau\right),
\nonumber\\
&&\V^{\tau\d}_{\rm sd}=-\frac{\sqrt2}{8}r_0\M_{2,1}^\tau+\frac{\sqrt2}{12}r_{1}\left(\M_{2,0}^\tau-\M_{0,0}^\tau\right)
-\frac{1}{4}\left(r_{1}\E_{1,0}^\tau-\frac{1}{\sqrt2}r_{0}\E_{1,1}^\tau\right)
+\frac{\sqrt2}{2}r_{-1}\M_{2,2}^\tau,
\nonumber\\
&&\V^{\tau\u}_{\rm sd}=\frac{\sqrt2}{8}r_0\M_{2,-1}^\tau-\frac{\sqrt2}{12}r_{-1}\left(\M_{2,0}^\tau-\M_{0,0}^\tau\right)
-\frac{1}{4}\left(r_{-1}\E_{1,0}^\tau+\frac{1}{\sqrt2}r_{0}\E_{1,-1}^\tau\right)
-\frac{\sqrt2}{2}r_{1}\M_{2,-2}^\tau.
\end{eqnarray}
%
with denotations
\begin{eqnarray}
&&\M_{2,1}\equiv R_1^--\sqrt{2}R_0^\d,\quad
  \M_{2,-1}\equiv R_{-1}^-+\sqrt{2}R_0^\u,\quad\M_{2,0}\equiv R_0^-+\left(R_1^\u-R_{-1}^\d\right)/\sqrt{2},\nonumber\\
&&\M_{2,2}\equiv -R_{1}^\d/\sqrt{2}\quad
  \M_{2,-2}\equiv R_{-1}^\u /\sqrt{2},\quad
  \M_{0,0}\equiv R_{0}^- - \sqrt{2}\left(R_1^\u-R_{-1}^\d\right)
  \nonumber\\
&&\E_{1,1}\equiv R_1^-+\sqrt2R_0^\d,\quad
\E_{1,-1}\equiv R_{-1}^--\sqrt2R_0^\u,\quad\E_{1,0}\equiv R_1^\u+R_{-1}^\d.
\nonumber
\end{eqnarray}
The contribution to mean field coming from
$\di\sum_{\mu}(-1)^{\mu}s_{1-\mu} s_{1\mu}$ is denoted as $\E_{1,\mu}$, the contribution from $\di\sum_{\mu}(-1)^{\mu}s_{2-\mu} s_{2\mu}$ is denoted as $\M_{2,\mu}$, and the contribution from $s_{00} s_{00}$ is denoted as $\M_{0,0}$, where $s_{\lambda\mu}$ is defined by eq.~(\ref{sd}).
Taking into account that
\begin{eqnarray}
&&\M_{2,1}-\E_{1,1}=-2\sqrt2R^\d_0,\quad\M_{2,-1}-\E_{1,-1}=2\sqrt2R^\u_0,\nonumber\\
&&\M_{2,1}+\E_{1,1}=2R^-_1,\quad\qquad\ \M_{2,-1}+\E_{1,-1}=2R^-_{-1},
\nonumber\\
&&\M_{2,0}-\M_{0,0}=\frac{3}{\sqrt2}\left(R_1^\u-R_{-1}^\d\right),\quad
\M_{2,0}+\frac{1}{2}\M_{0,0}=\frac{3}{2}R_0^-,\nonumber
\end{eqnarray}
we find from eq.~(\ref{Vsd})
%
\begin{equation}
 \V^{\tau\zeta}_{\rm sd}=\frac{1}{2}\left[
r_0R_0^{\tau\zeta}-r_{-1}R_1^{\tau\zeta}-r_1R^{\tau\zeta}_{-1}\right].
\end{equation}
 Finally we get:
 \begin{eqnarray}
  h^{\tau+}=V^{\tau+}_{\rm qq}+V^{\tau+}_{\rm dd},\quad
  h^{\tau\zeta}=V^{\tau\zeta}_{\rm sd}.
 \end{eqnarray}


 \section{Equations}
\label{AppEq}

\subsection{$K^\pi= 2^-$ equations}

\begin{eqnarray}
\dot {\bar\R}^\d_1&=&\frac{1}{m}\bar\P^\d_1
\nonumber\\
 \dot {\bar\P}^\d_1&=&
-\left(m\omega^2-2\kappa_0 Q_{20}\right)\bar\R^\d_1  +  2\alpha\kappa_0 \bar Q_{20}\R^\d_1
-\frac{A}{4}\left(\chi_1 \bar\R^\d_1 + \chi_0 X\R^\d_1\right)
,
\nonumber\\
\dot\R^\d_1&=&\frac{1}{m}\P^\d_1
\nonumber\\
 \dot\P^\d_1&=&
-\left(m\omega^2-2\kappa_0 Q_{20}\right)\R^\d_1  +  2\alpha\kappa_0 \bar Q_{20}\bar\R^\d_1
-\frac{A}{4}\left(\chi_0\R^\d_1 + \chi_1 X\bar\R^\d_1\right)
.
\label{mu=2(4)}
\end{eqnarray}
where $\bar Q_{20}=XQ_{20}$, $X=(N-Z)/A$,
$\di Q_{20}=Q_{00}\frac{4}{3}\delta$,
 $Q_{00}=A\langle r^2\rangle$,
$\di\kappa_0=-\frac{m\tilde\omega^2}{4Q_{00}}$~\cite{BaSc},
$\omega^2=\omega_0^2/[(1+\frac{4}{3}\delta)^{2/3}(1-\frac{2}{3}\delta)^{1/3}]$  with $\hbar\omega_0=41A^{-1/3}$. Notations $m\omega^2C_1=m\omega^2-2\kappa_0 Q_{20}$ and
$m\omega^2C_0=m\omega^2+4\kappa_0 Q_{20}$ are used in the main text of the paper, see section~\ref{GDR0}.

\subsection{$K^\pi= 1^-$ equations}

\begin{eqnarray}
\dot {\bar\R}^+_1&=&
\frac{1}{m}\bar\P^+_1  - i\hbar\frac{\eta}{2}\left(\bar\R^-_1+\sqrt2\bar\R^\d_0\right),
\nonumber \\
\dot {\bar\R}^-_1&=&
\frac{1}{m}\bar\P^-_1  - i\hbar\frac{\eta}{2}\bar\R^+_1,
\nonumber \\
\dot {\bar\R}^\d_0&=&
\frac{1}{m}\bar\P^\d_0  - i\hbar\frac{\eta}{2\sqrt2}\bar\R^+_1,
\nonumber \\
\dot {\bar\P}^+_1&=&
-\left(m\omega^2-2\kappa_0 Q_{20}\right)\bar\R^+_1  - i\hbar\frac{\eta}{2}\left(\bar\P^-_1+\sqrt2\bar\P^\d_0\right) + 2\alpha\kappa_0 \bar Q_{20}\R^+_1
-A\left(\xi_1^{K=1} \bar \R^+_1 + \xi_0^{K=1} X\R^+_1\right),
\nonumber \\
\dot {\bar\P}^-_1&=&
-\left(m\omega^2-2\kappa_0 Q_{20}\right)\bar\R^-_1  - i\hbar\frac{\eta}{2}\bar\P^+_1 + 2\alpha\kappa_0 \bar Q_{20}\R^-_1
-\frac{A}{4}\left(\chi_1 \bar\R^-_1 + \chi_0 X\R^-_1\right)
\nonumber \\
\dot {\bar\P}^\d_0&=&
-\left(m\omega^2+4\kappa_0 Q_{20}\right)\bar\R^\d_0  - i\hbar\frac{\eta}{2\sqrt2}\bar\P^+_1 - 4\alpha\kappa_0 \bar Q_{20}\R^\d_0
-\frac{A}{4}\left(\chi_1 \bar\R^\d_0 + \chi_0 X\R^\d_0\right),
\nonumber \\
\dot {\R}^+_1&=&
\frac{1}{m}\P^+_1  - i\hbar\frac{\eta}{2}\left(\R^-_1+\sqrt2\R^\d_0\right),
\nonumber \\
\dot {\R}^-_1&=&
\frac{1}{m}\P^-_1  - i\hbar\frac{\eta}{2}\R^+_1,
\nonumber \\
\dot {\R}^\d_0&=&
\frac{1}{m}\P^\d_0  - i\hbar\frac{\eta}{2\sqrt2}\R^+_1,
\nonumber \\
\dot {\P}^+_1&=&
-\left(m\omega^2-2\kappa_0 Q_{20}\right)\R^+_1  - i\hbar\frac{\eta}{2}\left(\P^-_1+\sqrt2\P^\d_0\right) + 2\alpha\kappa_0 \bar Q_{20}\bar\R^+_1
-A\left(\xi_0^{K=1}  \R^+_1 + \xi_1^{K=1} X\bar\R^+_1\right),
\nonumber \\
\dot {\P}^-_1&=&
-\left(m\omega^2-2\kappa_0 Q_{20}\right)\R^-_1  - i\hbar\frac{\eta}{2}\P^+_1 + 2\alpha\kappa_0 \bar Q_{20}\bar\R^-_1
-\frac{A}{4}\left(\chi_0 \R^-_1 + \chi_1 X\bar\R^-_1\right),
\nonumber \\
\dot {\P}^\d_0&=&
-\left(m\omega^2+4\kappa_0 Q_{20}\right)\R^\d_0  - i\hbar\frac{\eta}{2\sqrt2}\P^+_1 - 4\alpha\kappa_0 \bar Q_{20}\bar\R^\d_0
-\frac{A}{4}\left(\chi_0 \R^\d_0 + \chi_1 X\bar\R^\d_0\right).
\label{mu=1(12)}\end{eqnarray}

where $\hbar^2\eta=2\hbar\omega_0\kappa_{\rm N}$~\cite{Ring,Solov} and $\kappa_{\rm N}$
is Nilsson spin-orbit strength constant.

\subsection{$K^\pi=0^-$ equations}
\label{K0eq_16}

\begin{eqnarray}
\dot {\bar\R}^+_0&=&
\frac{1}{m}\bar\P^+_0  - i\hbar\frac{\eta}{2}\sqrt2\left(\bar\R^\u_1+\bar\R^\d_{-1}\right),
\nonumber \\
\dot {\bar\R}^-_0&=&
\frac{1}{m}\bar\P^-_0,
\nonumber \\
\dot {\bar\R}^\u_1&=&
\frac{1}{m}\bar\P^\u_1  - i\hbar\frac{\eta}{2\sqrt2}\bar\R^+_0,
\nonumber \\
\dot {\bar\R}^\d_{-1}&=&
\frac{1}{m}\bar\P^\d_{-1}  - i\hbar\frac{\eta}{2\sqrt2}\bar\R^+_0,
\nonumber \\
\dot {\bar\P}^+_0&=&
-\left(m\omega^2+4\kappa_0 Q_{20}\right)\bar\R^+_0  - i\hbar\frac{\eta}{2}\sqrt2\left(\bar\P^\u_1+\bar\P^\d_{-1}\right) -4\alpha\kappa_0 \bar Q_{20}\R^+_0 -A\left(\xi_1^{K=0} \bar \R^+_0 + \xi_0^{K=0} X\R^+_0\right),
\nonumber\\
\dot {\bar\P}^-_0&=&
-\left(m\omega^2+4\kappa_0 Q_{20}\right)\bar\R^-_0  -4\alpha\kappa_0 \bar Q_{20}\R^-_0
-\frac{A}{4}\left(\chi_1\bar\R_0^- + \chi_0 X\R_0^-\right),
\nonumber \\
\dot {\bar\P}^\u_1&=&
-\left(m\omega^2-2\kappa_0 Q_{20}\right)\bar\R^\u_1  - i\hbar\frac{\eta}{2\sqrt2}\bar\P^+_0 + 2\alpha\kappa_0 \bar Q_{20}\R^\u_1
-\frac{A}{4}\left(\chi_1 \bar\R_1^\u + \chi_0 X\R_1^\u\right)
,
\nonumber \\
\dot {\bar\P}^\d_{-1}&=&
-\left(m\omega^2-2\kappa_0 Q_{20}\right)\bar\R^\d_{-1}  - i\hbar\frac{\eta}{2\sqrt2}\bar\P^+_0 + 2\alpha\kappa_0 \bar Q_{20}\R^\d_{-1}
-\frac{A}{4}\left(\chi_1 \bar\R_{-1}^\d + \chi_0 X\R_{-1}^\u\right)
,
\nonumber \\
\dot {\R}^+_0&=&
\frac{1}{m}\P^+_0  - i\hbar\frac{\eta}{2}\sqrt2\left(\R^\u_1+\R^\d_{-1}\right),
\nonumber \\
\dot {\R}^-_0&=&
\frac{1}{m}\P^-_0,
\nonumber\\
\dot {\R}^\u_1&=&
\frac{1}{m}\P^\u_1  - i\hbar\frac{\eta}{2\sqrt2}\R^+_0,
\nonumber \\
\dot {\R}^\d_{-1}&=&
\frac{1}{m}\P^\d_{-1}  - i\hbar\frac{\eta}{2\sqrt2}\R^+_0,
\nonumber \\
\dot {\P}^+_0&=&
-\left(m\omega^2+4\kappa_0 Q_{20}\right)\R^+_0  - i\hbar\frac{\eta}{2}\sqrt2\left(\P^\u_1+\P^\d_{-1}\right) -4\alpha\kappa_0 \bar Q_{20}\bar\R^+_0
-A\left(\xi_0^{K=0} \R^+_0 + \xi_1^{K=0} X \bar\R^+_0\right),
\nonumber \\
\dot {\P}^-_0&=&
-\left(m\omega^2+4\kappa_0 Q_{20}\right)\R^-_0  -4\alpha\kappa_0 \bar Q_{20}\bar\R^-_0
-\frac{A}{4}\left(\chi_0\R_0^- + \chi_1 X\bar\R_0^-\right)
\nonumber \\
\dot {\P}^\u_1&=&
-\left(m\omega^2-2\kappa_0 Q_{20}\right)\R^\u_1  - i\hbar\frac{\eta}{2\sqrt2}\P^+_0 + 2\alpha\kappa_0 \bar Q_{20}\bar\R^\u_1
-\frac{A}{4}\left(\chi_0 \R_1^\u + \chi_1 X\bar\R_1^\u\right)
,
\nonumber \\
\dot {\P}^\d_{-1}&=&
-\left(m\omega^2-2\kappa_0 Q_{20}\right)\R^\d_{-1}  - i\hbar\frac{\eta}{2\sqrt2}\P^+_0 + 2\alpha\kappa_0 \bar Q_{20}\bar\R^\d_{-1}
-\frac{A}{4}\left(\chi_0 \R_{-1}^\d + \chi_1 X\bar\R_{-1}^\u\right)
.
\label{mu=0(16)}\end{eqnarray}

\newpage
\subsubsection{Electric, $K^\pi=0^-$}
\label{K0eq_E}

\begin{eqnarray}
\dot {\bar\R}^+_0&=&
\frac{1}{m}\bar\P^+_0  - i\hbar\frac{\eta}{\sqrt2}\bar\R^E_0,
\nonumber \\
\dot {\bar\R}^E_0&=&
\frac{1}{m}\bar\P^E_0  - i\hbar\frac{\eta}{\sqrt2}\bar\R^+_0,
\nonumber \\
\dot {\bar\P}^+_0&=&
-m\omega^2C_0\bar\R^+_0  - i\hbar\frac{\eta}{\sqrt2}\bar P^E_0
+ \frac{4}{3}\alpha\delta X m\tilde\omega^2\R^+_0
-A\left(\xi_1^{K=0} \bar \R^+_0 + \xi_0^{K=0} X\R^+_0\right),
\nonumber\\
\dot {\bar\P}^E_0&=&
-m\omega^2C_1\bar\R^E_0  - i\hbar\frac{\eta}{\sqrt2}\bar\P^+_0
- \frac{2}{3}\alpha\delta X m\tilde\omega^2\R^E_0
-\frac{A}{4}\left(\chi_1 \bar\R^E_0 + \chi_0 X\R^E_0\right)
,
\nonumber \\
\dot {\R}^+_0&=&
\frac{1}{m}\P^+_0  - i\hbar\frac{\eta}{\sqrt2}\R^E_0,
\nonumber \\
\dot {\R}^E_0&=&
\frac{1}{m}\P^E_0  - i\hbar\frac{\eta}{\sqrt2}\R^+_0,
\nonumber \\
\dot {\P}^+_0&=&
-m\omega^2C_0\R^+_0  - i\hbar\frac{\eta}{\sqrt2}
\P^E_0 + \frac{4}{3}\alpha\delta X m\tilde\omega^2\bar\R^+_0
-A\left(\xi_0^{K=0} \R^+_0 + \xi_1^{K=0} X \bar\R^+_0\right),
\nonumber \\
\dot {\P}^E_0&=&
-m\omega^2C_1\R^E_0  - i\hbar\frac{\eta}{\sqrt2}\P^+_0 - \frac{2}{3}\alpha\delta X m\tilde\omega^2\bar\R^E_0
-\frac{A}{4}\left(\chi_0 \R^E_0+ \chi_1 X\bar R^E_0\right)
,
\label{mu=0(E)}
\end{eqnarray}
where $\R^E_0=\R_1^\u+\R_{-1}^\d$, $\P^E_0=\P_1^\u+\P_{-1}^\d$.

\subsubsection{Magnetic, $K^\pi=0^-$}
\label{K0eq_M}

\begin{eqnarray}
\dot {\bar\R}^-_0&=&
\frac{1}{m}\bar\P^-_0,
\nonumber\\
\dot {\bar\P}^-_0&=&
-m\omega^2C_0\bar\R^-_0  + \frac{4}{3}\alpha\delta X m\tilde\omega^2\R^-_0
-\frac{A}{4}\left(\chi_1\bar\R_0^- + \chi_0 X\R_0^-\right),
\nonumber \\
\dot {\R}^-_0&=&
\frac{1}{m}\P^-_0,
\nonumber\\
\dot {\P}^-_0&=&
-m\omega^2C_0\R^-_0  + \frac{4}{3}\alpha\delta X m\tilde\omega^2\bar\R^-_0
-\frac{A}{4}\left(\chi_0\R_0^- + \chi_1 X\bar\R_0^-\right),
\label{mu=0(M-SV)}\end{eqnarray}
\begin{eqnarray}
\dot {\bar\R}^M_0&=&
\frac{1}{m}\bar\P^M_0,
\nonumber \\
\dot {\bar\P}^M_0&=&
-m\omega^2C_1\bar\R^M_0 - \frac{2}{3}\alpha\delta X m\tilde\omega^2\R^M_0
-\frac{A}{4}\left(\chi_1\bar\R_0^M + \chi_0 X\R_0^M\right)
,
\nonumber \\
\dot {\R}^M_0&=&
\frac{1}{m}\P^M_0,
\nonumber \\
\dot {\P}^M_0&=&
-m\omega^2C_1\R^M_0 - \frac{2}{3}\alpha\delta X m\tilde\omega^2\bar\R^M_0
-\frac{A}{4}\left(\chi_0\R_0^M + \chi_1 X\bar\R_0^M\right)
,
\label{mu=0(M-SF)}\end{eqnarray}
where $\R^M_0=\R_1^\u-\R_{-1}^\d$, $\P^M_0=\P_1^\u-\P_{-1}^\d$.

\newpage
\section{Electromagnetic transition probabilities}
\label{AppEM}

 Excitation
probabilities are calculated with the help of the theory of linear
response of the system to a weak external field
\begin{equation}
\label{extf}
\hat O(t)=\hat O \,\e^{-i\Omega t}+\hat O^{\dagger}\,e^{i\Omega t}.
\end{equation}
A detailed explanation can be found in \cite{BaSc,BaMoPRC22}.
We recall only the main points.
The matrix elements of the operator $\hat O$ obey the relationship \cite{Lane}
\begin{equation}
\label{matel}
|\langle\psi_a|\hat O|\psi_0\rangle|^2=
\hbar\lim_{\Omega\to\Omega_a}(\Omega-\Omega_a)
\overline{\langle\psi'|\hat O|\psi'\rangle\e^{-i\Omega t}},
\end{equation}
where $\psi_0$ and $\psi_a$ are the stationary wave functions of the
unperturbed ground and excited states; $\psi'$ is the wave function
of the perturbed ground state, $\Omega_a=(E_a-E_0)/\hbar$ are the
normal frequencies, the bar means averaging over a time interval much
larger than $1/\Omega$.

\subsection{Electric dipole transitions}
\label{AppA}
To calculate the electric dipole transition probability, it is necessary
to excite the system by the following external field:
\begin{equation}
\label{OGDR}
\hat O(E1,\mu)=e\sum_\tau\left(\frac{Z}{A}\delta_{\tau{\rm n}}-\frac{N}{A}\delta_{\tau{\rm p}}\right)
r_\tau Y_{1\mu}(\hat\br_\tau).
\end{equation}
The electric dipole operator (\ref{OGDR})
in cyclic coordinates $r_\mu$
\begin{equation}
\label{OGDR_1}
\hat O(E1,\mu)=e\sqrt{\frac{3}{4\pi}}\sum_\tau\left(\frac{Z}{A}\delta_{\tau{\rm n}}-\frac{N}{A}\delta_{\tau{\rm p}}\right)
[r_\mu]_\tau.
\end{equation}
The  reduced  probability  for  electric  dipole  transition reads
\begin{equation}\label{BE1}
B(E1\mu)_a=|\langle\psi_a|\hat O(E1,\mu)|\psi_0\rangle|^2=\hbar\lim_{\Omega\to\Omega_a}(\Omega-\Omega_a)
\overline{\langle\psi'|\hat O(E1,\mu)|\psi'\rangle\e^{-i\Omega t}}.
\end{equation}
For the matrix element we have
\begin{eqnarray}
\label{OE1}
\langle\psi'|\hat O(E1,\mu)|\psi'\rangle =
e\sqrt{\frac{3}{4\pi}}\left[\frac{Z}{A}\R_\mu^{\rm n}-\frac{N}{A}\R_\mu^{\rm p}\right]=
\frac{e}{2}\sqrt{\frac{3}{4\pi}}\left[\bar\R_\mu^+-X\R_\mu^+\right].
\end{eqnarray}
Due to the external field (\ref{OGDR_1}) dynamical equations for collective variables $\P_\mu^{{\rm n}+}$ and $\P_\mu^{{\rm p}+}$
become inhomogeneous:
\begin{eqnarray}
     \dot\P_\mu^{{\rm n}+}
&=&\ldots\; +e\sqrt{\frac{3}{4\pi}}\frac{NZ}{A}\,\e^{i\Omega t},
\nonumber\\
     \dot\P_\mu^{{\rm p}+}
&=&\ldots\; -e\sqrt{\frac{3}{4\pi}}\frac{NZ}{A}\,\e^{i\Omega t}.
\end{eqnarray}
As a result for $\bar\P_\mu^{+}=\P_\mu^{{\rm n}+}-\P_\mu^{{\rm p}+}$  we obtain
\begin{equation}
     \dot{\bar\P}_\mu^{+}
=\ldots\; +2e\sqrt{\frac{3}{4\pi}}\frac{NZ}{A}\,\e^{i\Omega t}.
\end{equation}
Solving the inhomogeneous set of equations
one can find the required
in (\ref{BE1}), (\ref{OE1})
values of $\bar\R_\mu^+$ and $\R_\mu^+$
 and  calculate
$B(E1\mu)$ factors for all excitations as it is explained in \cite{BaSc,BaMoPRC22}.


\subsection{Magnetic transitions}
\label{AppB}

To calculate the magnetic transition probability, it is necessary
to excite the system by the following external field:
\begin{equation}
\label{OM2}
\hat O(M\lambda,\mu)=\mu_N\sum\limits_{k}^{A}\left(
g_s(k)\hat{\bf S}_k+\frac{2}{\lambda+1}g_l(k)\hat{\bf l}_k
\right)\cdot\nabla_k \left(r_k^\lambda Y_{\lambda\mu}(\hat\br_k)\right).
\end{equation}
The free particle $g$-factors are given by
$g_l^{\rm p}=1,$ $g_s^{\rm p}=5.5856$ for protons and $g_l^{\rm n}=0,$ $g_s^{\rm n}=-3.8263$ for neutrons,
$g_s^{\rm IV}=g_s^{\rm n}-g_s^{\rm p}$, $g_s^{\rm IS}=g_s^{\rm n}+g_s^{\rm p}$.
The spin quenching factor $q$ can be applied in the calculations: $g_s^\tau=q g_s^{\tau\rm free}$.
The  reduced  probability  for $M\lambda$ magnetic  transition reads
\begin{equation}
B(M\lambda\mu)_a=|\langle\psi_a|\hat O(M\lambda,\mu)|\psi_0\rangle|^2
=\hbar\lim_{\Omega\to\Omega_a}(\Omega-\Omega_a)
\overline{\langle\psi'|\hat O(M\lambda,\mu)|\psi'\rangle\e^{-i\Omega t}}.
\end{equation}
The $\hat O(M2,\mu)$ operator in cyclic coordinates reads
\begin{equation}
\label{OM2sl}
\hat O(M2,\mu)=\mu_N\sqrt{\frac{15}{2\pi}}\sum\limits_{k}^{A}
\sum\limits_{\nu=-1}^{1}C_{1(\mu-\nu),1\nu}^{2\mu}\left[g_s\hat S_\nu r_{\mu-\nu}+\frac{2}{3}g_l\hat l_\nu r_{\mu-\nu}\right]_k.
\end{equation}
For the matrix element we have
\begin{equation}
\label{M2sl}
\langle\psi'|\hat O(M2,\mu)|\psi'\rangle =
\mu_N\sqrt{\frac{15}{2\pi}}\sum\limits_{k}^{A}
\sum\limits_{\nu=-1}^{1}C_{1(\mu-\nu),1\nu}^{2\mu}\left[g_s\sum_{\sigma,\sigma'}\langle\sigma|\hat S_\nu|\sigma'\rangle
\R^{\sigma'\sigma}_{\mu-\nu}
-i\frac{2\sqrt2}{3} g_l\T^+_{1\nu,\mu-\nu}
\right]_k,
\end{equation}
where
\begin{equation}
\di\T^+_{\lambda\mu,\nu}(t)=(2\pi\hbar)^{-3}\int\! d\br\int\! d\bp\,
\{r\otimes p\}_{\lambda\mu} r_\nu
\,\delta f^+(\br,\bp,t).
\label{T3}
\end{equation}
Only the dynamics of tensors of the first rank in the coordinate and momentum are taken into account in this work.
In this formulation of the problem, only the spin contribution of the operator $\hat O(M2,\mu)$ can be estimated:
\begin{equation}
\label{M2s}
\langle\psi'|\hat O_S(M2,\mu)|\psi'\rangle =
\mu_N\sqrt{\frac{15}{2\pi}}\sum\limits_{\tau={\rm n},{\rm p}} g_s^\tau
\sum\limits_{\nu=-1}^{1}C_{1(\mu-\nu),1\nu}^{2\mu}\left[\sum_{\sigma,\sigma'}\langle\sigma|\hat S_\nu|\sigma'\rangle
\R^{\sigma'\sigma}_{\mu-\nu}\right]_\tau.
\end{equation}
From eq.~(\ref{M2s}) for the $\mu=0, 1, 2$
components we obtain
\begin{eqnarray}
\label{Ms20}
&&\langle\psi'|\hat O_S(M2,0)|\psi'\rangle =
\mu_N\sqrt{\frac{5}{4\pi}}\hbar\sum\limits_{\tau={\rm n},{\rm p}}
g_s^{\tau}\left[\R_0^{\tau-} +\frac{1}{\sqrt2}\left(\R_1^{\tau\u}-\R_{-1}^{\tau\d}\right)\right],
\\
\label{Ms21}
&&\langle\psi'|\hat O_S(M2,1)|\psi'\rangle =
\mu_N\sqrt{\frac{15}{16\pi}}\hbar\sum\limits_{\tau={\rm n},{\rm p}}
g_s^{\tau}\left[\R_1^{\tau-} -\sqrt2\,\R_0^{\tau\d}\right],\\
\label{Ms22}
&&\langle\psi'|\hat O_S(M2,2)|\psi'\rangle =-
\mu_N\sqrt{\frac{15}{4\pi}}\hbar\sum\limits_{\tau={\rm n},{\rm p}}
g_s^{\tau}\R_1^{\tau\d}.
\end{eqnarray}
The matrix elements $\langle\psi'|\hat O_S(M2,\mu)|\psi'\rangle$
in terms of isoscalar and isovector variables have the following form
\begin{eqnarray}
\label{M20}
&&\langle\psi'|\hat O_S(M2,0)|\psi'\rangle =
\mu_N\frac{\hbar}{2}\sqrt{\frac{5}{4\pi}}
\left[g_s^{\rm IS}\left(\R_0^- +\frac{\R_1^\u-\R_{-1}^\d}{\sqrt2}\right)+
g_s^{\rm IV}\left(\bar\R_0^- +\frac{\bar\R_1^\u-\bar\R_{-1}^\d}{\sqrt2}\right)\right], \\
\label{M21}
&&\langle\psi'|\hat O_S(M2,1)|\psi'\rangle =
\mu_N\frac{\hbar}{2}\sqrt{\frac{15}{16\pi}}
\left[g_s^{\rm IS}\left(\R_1^- -\sqrt2\,\R_0^\d\right)+
g_s^{\rm IV}\left(\bar\R_1^- -\sqrt2\,\bar\R_0^\d\right)\right], \\
\label{M22}
&&\langle\psi'|\hat O_S(M2,2)|\psi'\rangle =-
\mu_N\frac{\hbar}{2}\sqrt{\frac{15}{4\pi}}
\left[g_s^{\rm IS}\R_1^\d+g_s^{\rm IV}\bar\R_1^\d\right].
\end{eqnarray}
Since $g_s^{\rm IV}=9.4119$ is much larger than $g_s^{\rm IS}=1.7593$ the magnetic multipole operator is dominantly isovector in nature.

Due to the external field (\ref{OM2sl}), some dynamical equations become inhomogeneous:
\begin{eqnarray}
\dot\P_1^{{\tau}-}
&=&\ldots\; +\mu_N\hbar\sqrt{\frac{15}{4\pi}} \frac{g_s^{\tau}}{2}N^\tau\,\e^{i\Omega t},
\nonumber\\
\dot\P_0^{{\tau}\d}
&=&\ldots\; -\mu_N\hbar\sqrt{\frac{15}{8\pi}} \frac{g_s^{\tau}}{2}N^\tau\,\e^{i\Omega t},
\nonumber\\
\dot\P_1^{{\tau}\d}
&=&\ldots\; +\mu_N\hbar\sqrt{\frac{15}{4\pi}} \frac{g_s^{\tau}}{2}N^\tau
\,\e^{i\Omega t},
\nonumber\\
     \dot\P_0^{{\tau}-}
&=&\ldots\; -\mu_N\hbar\sqrt{\frac{5}{4\pi}} g_s^{\tau} N^\tau
\,\e^{i\Omega t},
\nonumber\\
\dot\P^{\tau\u}_1&=&
\ldots\; -\mu_N\hbar\sqrt{\frac{5}{8\pi}} \frac{g_s^{\tau}}{2}N^\tau
\,\e^{i\Omega t},
\nonumber\\
\dot \P^{\tau\d}_{-1}&=&
\ldots\; +\mu_N\hbar\sqrt{\frac{5}{8\pi}} \frac{g_s^{\tau}}{2}N^\tau
\,\e^{i\Omega t},
\end{eqnarray}
where $N^\tau=N$ for $\tau={\rm n}$ and $N^\tau=Z$ for $\tau={\rm p}$.
As a result for isovector and isoscalar  equations from (\ref{mu=2(4)}), (\ref{mu=1(12)}) and (\ref{mu=0(16)})  we obtain
\begin{eqnarray}
\label{inhM}
     \dot{\bar\P}_1^{-}
&=&\ldots\; +\mu_N\hbar\sqrt{\frac{15}{4\pi}}\frac{A}{4}\left( g_s^{\rm IV} +Xg_s^{\rm IS} \right)\,\e^{i\Omega t},
\nonumber\\
     \dot\P_1^{-}
&=&\ldots\; +\mu_N\hbar\sqrt{\frac{15}{4\pi}}\frac{A}{4}\left( g_s^{\rm IS} +Xg_s^{\rm IV} \right)\,\e^{i\Omega t},
\nonumber\\
\label{inhMx}
     \dot{\bar\P}_0^\d
&=&\ldots\; -\mu_N\hbar\sqrt{\frac{15}{8\pi}}\frac{A}{4}\left( g_s^{\rm IV} +Xg_s^{\rm IS} \right)\,\e^{i\Omega t},
\nonumber\\
     \dot\P_0^\d
&=&\ldots\; -\mu_N\hbar\sqrt{\frac{15}{8\pi}}\frac{A}{4}\left( g_s^{\rm IS} +Xg_s^{\rm IV} \right)\,\e^{i\Omega t},
\nonumber\\  \label{inhM0}
\dot {\bar\P}^\d_1&=&
\ldots\; +\mu_N\hbar\sqrt{\frac{15}{4\pi}}\frac{A}{4}\left(g_s^{\rm IV}+Xg_s^{\rm IS}\right)
\,\e^{i\Omega t},
\nonumber\\
\dot {\P}^\d_1&=&
\ldots\; +\mu_N\hbar\sqrt{\frac{15}{4\pi}}\frac{A}{4}\left(g_s^{\rm IS}+Xg_s^{\rm IV}\right)
\,\e^{i\Omega t}, \nonumber\\
     \dot{\bar\P}_0^{-}
&=&\ldots\; -\mu_N\hbar\sqrt{\frac{5}{4\pi}}\frac{A}{2}\left( g_s^{\rm IV} +Xg_s^{\rm IS} \right)\, \,\e^{i\Omega t},
\nonumber\\
     \dot\P_0^{-}
&=&\ldots\; -\mu_N\hbar\sqrt{\frac{5}{4\pi}}\frac{A}{2}\left( g_s^{\rm IS} +Xg_s^{\rm IV} \right)\, \,\e^{i\Omega t},
\nonumber\\
\dot {\bar\P}^M_0&=&
\ldots\; -\mu_N\hbar\sqrt{\frac{5}{8\pi}}\frac{A}{2}\left(g_s^{\rm IV}+Xg_s^{\rm IS}\right)
\,\e^{i\Omega t},
\nonumber\\
\dot {\P}^M_0&=&
\ldots\; -\mu_N\hbar\sqrt{\frac{5}{8\pi}}\frac{A}{2}\left(g_s^{\rm IS}+Xg_s^{\rm IV}\right)
\,\e^{i\Omega t}.
\end{eqnarray}
Solving the inhomogeneous sets of equations (\ref{inhM})
one can find the required in  (\ref{M20}), (\ref{M21}), (\ref{M22}) values of $\bar\R_\mu^\zeta$ and $\R_\mu^\zeta$ ($\zeta=-,\u,\d$)
 and  calculate $B(M2\mu)$ factors.
\end{widetext}

\end{document}